\documentclass[12pt]{article} 
\usepackage[sectionbib]{natbib}
\usepackage{array,epsfig,fancyheadings,rotating}
\usepackage{sectsty, secdot}
\sectionfont{\fontsize{12}{14pt plus.8pt minus .6pt}\selectfont}
\renewcommand{\theequation}{\thesection\arabic{equation}}
\subsectionfont{\fontsize{12}{14pt plus.8pt minus .6pt}\selectfont}

\usepackage[
top    = 1in,
bottom = 1in,
left   = 1in,
right  = 1in]{geometry}

\usepackage{mathrsfs}
\usepackage{multirow}
\usepackage{algorithm}
\usepackage[noend]{algorithmic}
\usepackage[
    colorlinks=true,%
    breaklinks,%
    hyperindex,%
    plainpages=false,%
    bookmarksopen=false,%
    linkcolor=blue,%
    anchorcolor=blue,%
    citecolor=blue,%
    filecolor=black,%
    menucolor=black,%
    pagecolor=black,%
    urlcolor=blue,%
    pdfview=FitH,%
    pdfstartview=FitH,%
    hyperfigures,
    bookmarksnumbered%
  ]{hyperref}
\usepackage{amsmath,amssymb,amsthm,amsfonts,amsbsy,latexsym,dsfont,mathtools,enumerate,graphicx,setspace}

\newtheorem{thm}{Theorem}
\newtheorem{lem}{Lemma}
\newtheorem{cor}{Corollary}
\newtheorem{prop}{Proposition}
\theoremstyle{definition}

\newtheorem{ex}{Example}
\newtheorem{rem}{Remark}
\makeatletter
\newcommand{\oset}[3][0ex]{%
  \mathrel{\mathop{#3}\limits^{
    \vbox to#1{\kern-2\ex@
    \hbox{$\scriptstyle#2$}\vss}}}}
\makeatother

\newcommand{\Cov}[0]{\text{Cov}}
\newcommand{\Var}[0]{\text{Var}}

\newcommand{\SP}[0]{\text{SP}}
\newcommand{\GP}[0]{\text{GP}}
\newcommand{\Ber}[0]{\mathsf{Ber}}

\renewcommand{\leq}{\leqslant}
\renewcommand{\geq}{\geqslant}

\newcommand{\rC}{\mathrm{C}} \newcommand{\rS}{\mathrm{S}}

\newcommand{\cB}{\mathcal{B}}  
\newcommand{\cG}{\mathcal{G}} \newcommand{\cV}{\mathcal{V}}
\newcommand{\vone}{\mathbf{1}}

\newcommand{\vU}{\mathbf{U}}
\newcommand{\vX}{\mathbf{X}}\newcommand{\vY}{\mathbf{Y}}

\newcommand{\mvZ}{\boldsymbol{Z}}

\newcommand{\mvv}{\boldsymbol{v}}

\newcommand{\mvdelta}{\boldsymbol{\delta}}
\newcommand{\bG}{\mathbb{G}}
\newcommand{\bH}{\mathbb{H}}
\newcommand{\sF}{\mathscr{F}}
\newcommand{\rd}{\mathrm{d}}

\newcommand{\myfontsizeA}{\fontsize{10}{12}\selectfont}
\newcommand{\myfontsizeAA}{\fontsize{8}{11}\selectfont}
\DeclareMathOperator{\E}{\mathds{E}}

\DeclareMathOperator{\cov}{Cov}

\DeclareMathOperator{\argmin}{argmin}

\newcommand\myeq{\stackrel{\mathclap{\normalfont{d}}}{=}}
\newcommand\mysim{\stackrel{\mathclap{\normalfont{i.i.d.}}}{\sim}}
\newcommand\convD{\stackrel{\mathclap{\normalfont{d}}}{\rightarrow}}

\makeatletter
\newenvironment{breakablealgorithm}
  {
   \begin{center}
     \refstepcounter{algorithm}
     \hrule height.8pt depth0pt \kern2pt
     \renewcommand{\caption}[2][\relax]{
       {\raggedright\textbf{\ALG@name~\thealgorithm} ##2\par}%
       \ifx\relax##1\relax 
         \addcontentsline{loa}{algorithm}{\protect\numberline{\thealgorithm}##2}%
       \else 
         \addcontentsline{loa}{algorithm}{\protect\numberline{\thealgorithm}##1}%
       \fi
       \kern2pt\hrule\kern2pt
     }
  }{
     \kern2pt\hrule\relax
   \end{center}
  }
\makeatother

\begin{document}


\renewcommand{\baselinestretch}{2}


\markboth{\hfill{\footnotesize\rm Yeonjoo Park, Xiaohui Chen, and Douglas G. Simpson} \hfill}
{\hfill {\footnotesize\rm Robust Inference for Partially Observed Functional Response Data} \hfill}

\renewcommand{\thefootnote}{}
$\ $\par


\fontsize{12}{14pt plus.8pt minus .6pt}\selectfont \vspace{0.8pc}
\centerline{\large\bf Robust Inference for Partially Observed}
\vspace{2pt} \centerline{\large\bf Functional Response Data}
\vspace{.4cm} \centerline{Yeonjoo Park$^1$, Xiaohui Chen$^2$, and Douglas G. Simpson$^2$} \vspace{.4cm} \centerline{\it
$^1$University of Texas at San Antonio and}
\vspace{2pt} \centerline{\it $^2$University of Illinois at Urbana-Champaign}
 \vspace{.55cm} \fontsize{9}{11.5pt plus.8pt minus
.6pt}\selectfont


\begin{quotation}
\noindent {\it Abstract:}
Irregular functional data in which densely sampled curves are observed over different ranges pose a challenge for modeling and inference, and sensitivity to outlier curves is a concern in applications. Motivated by applications in quantitative ultrasound signal analysis, this paper investigates a class of robust M-estimators for partially observed functional data including functional location and quantile estimators. Consistency of the estimators is established under general conditions on the partial observation process. Under smoothness conditions on the class of M-estimators, asymptotic Gaussian process approximations are established and used for large sample inference. The large sample approximations justify a bootstrap approximation for robust inferences about the functional response process. The performance is demonstrated in simulations and in the analysis of irregular functional data from quantitative ultrasound analysis.

\vspace{9pt}
\par
\end{quotation}\par

\def\thefigure{\arabic{figure}}
\def\thetable{\arabic{table}}

\renewcommand{\theequation}{\thesection.\arabic{equation}}

\fontsize{12}{14pt plus.8pt minus .6pt}\selectfont

\setcounter{section}{0} 
\setcounter{equation}{0} 

\lhead[\footnotesize\thepage\fancyplain{}\leftmark]{}\rhead[]{\fancyplain{}\rightmark\footnotesize\thepage}

\newpage

\section{Introduction}

With advances in instrumentation and the capability to acquire data densely over a continuum, function-valued data acquisition is increasingly common in many fields; see, e.g. \citet*{ramsay2005} and  \citet*{horvath2012}.
Earlier works on functional data focused in large part on regular functional data, where the functional samples are densely collected over a common domain, or sparse functional data, in which the functional response for each subject is sparsely sampled with a small number of irregularly spaced measurements over the domain. In recent years, applications have emerged that produce partially observed functional data, where each individual trajectory is only collected over individual specific subinterval(s) densely or even sparsely within the whole domain of interest. Several recent works have begun addressing the estimation of covariance functions for short functional segments observed at sparse and irregular grid points, called ``functional snippets", (\citet*{descary2019}; \citet*{lin2020a}; \citet*{lin2020b}; \citet*{zhang2020}), or for fragmented functional data observed on small subintervals (\citet*{delaigle2020}).  For densely observed partial data, existing studies include the estimation of the unobserved part of curves (\citet*{kraus2015}; \citet*{ delaigle2016};  \citet*{kneip2019}), prediction (\citet*{liebl2013}; \citet*{goldberg2014}), classification (\citet*{delaigle2013}; \citet*{stefanucci2018}; \citet*{mojirsheibani2018}; \citet*{kraus2019A}; \citet*{parksimpson2019}), functional regression (\cite{gellar2014}), and inferences (\citet*{gromenko2017}; \citet*{kraus2019}).

Robustness to atypical curves or deviations from Gaussian variation is a practical concern in modeling and inference, especially for partially observed functional data. For example, \citet*{parksimpson2019} demonstrated that t-type heavy-tailed models for functional data performed better than Gaussian methods for probabilistic classification of quantitative ultrasound (QUS) measurements, which extract diagnostic information on biological tissues, such as tumors, from the ultrasound radio frequency backscattering signals. In QUS analysis the backscattered spectrum is captured by transducer by scanning the region of interest. The attenuation adjusted backscatter coefficient (BSC) comprises a  functional curve spanning the frequency range of the transducer.

\begin{figure} [t!]
  \centering
  \includegraphics[width=3.5in]{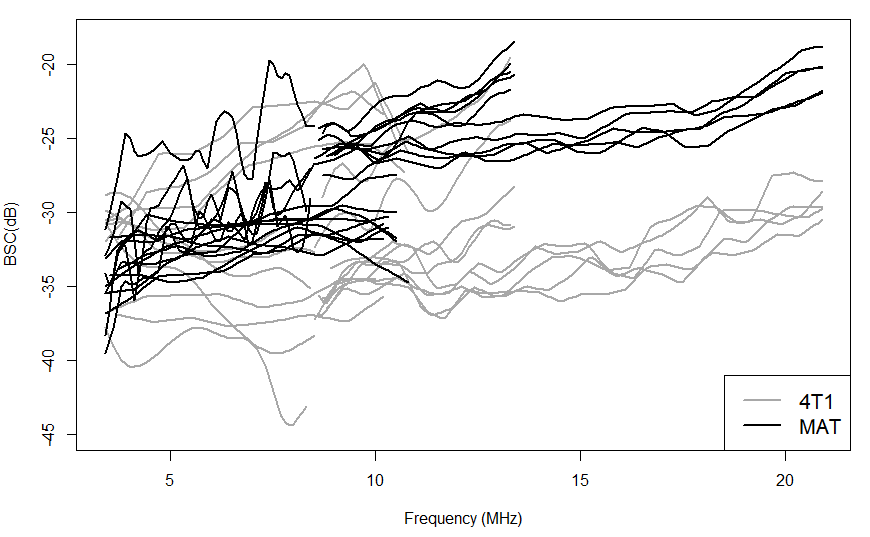}
  \caption{BSC data example by scanning two mammary tumors, 4T1 and MAT.}\label{MAT4T1_example}
\end{figure}

\cite{wirtzfeld2015} presented QUS data from an inter-laboratory diagnostic ultrasound study in which two types of induced mammary tumors were scanned using multiple transducers of varying center frequencies: 4T1 tumors in mice and MAT tumors in rats. Figure \ref{MAT4T1_example} shows a subset of the data. The resulting BSC curves are observed over varying frequency ranges depending on transducers used in scanning, and at the same time, several curves show atypical behaviors, especially at the lower frequency ranges in the 4T1 group.

The example illustrates issues that motivate two main goals in this article: (i) to develop a robust functional data analysis approach that is general enough to handle partially observed functional data; and (ii) to establish asymptotic properties to provide the foundation for associated robust inferences. 

Several authors have studied robust estimation for balanced functional data. Works by \citet*{fraiman2001}, \citet*{cuevas2007}, and \citet*{lopez-pintado2009, lopez-pintado2011} extended the data-depth notion in robust multivariate data analysis to functional data and defined depth weighted robust estimators. \cite{locatore1999}, \citet*{gervini2008} and \citet*{sinova2018} developed robust estimators from fully functional approach with investigations on robustness and asymptotic properties of the estimators. None of these methods is directly applicable to partially observed functional data.

We propose a new class of functional M-estimator by extending a class of M-estimators \citet*{huber2005} to functional data. The approach considered here is in contrast to the recent functional location M-estimators developed in \citet*{sinova2018}, which imposed a bounded M-estimator score function on the norm of the entire functional deviation from the location parameter function. Our approach builds the robust estimator in a cross-sectional manner to take advantage of all available curve data at each spatial location, while adapting to uneven patches in the response samples due to partial observation of individual curves. The difference will be presented in detail in Section \ref{sec:MMest}. Even with fully observed functional data, the cross-sectional approach considered here demonstrates a capability to adjust outlying patches in local spatial locations better than the via robust pseudo-norm to the entire function.

We employ a missing data approach to deal with the partially observed functional data, extending the  framework considered by \citet*{kraus2015, kraus2019} and \citet*{park2017} for functional mean and covariance estimation. We establish asymptotic properties of the M-estimator including the consistency and Gaussian process approximations. Furthermore, we adapt the results to develop a robust functional ANOVA test using $L^2$-norm statistics, and a functional trend test for shape inferences, in each case implementing the inferences via a bootstrap approach. The robustness of the M-estimators is investigated by influence analysis to establish the bounded influence of outlying curves. Simulation studies and real data analysis from a QUS study demonstrate the properties of the methods.

Section \ref{sec:MEST} defines the new class of functional M-estimators. The approach taken here has the advantage of being directly applicable to partially observed functional data while bounding the influence of extreme curves. Section \ref{sec:Theory} establishes the theoretical properties of M-estimator including the consistency and the Gaussian process approximations of the estimates. These results are then used to develop the $L^2$-norm test and functional trend test, and to support the bootstrap inferences for practical implementation. The remaining sections include the simulations and real data examples. Technical proofs and additional simulation results are included in the online supplementary material.

\section{M-estimator for Partially Observed Functional Data}
\label{sec:MEST}
\subsection{Modeling assumptions  }
\label{sec:model}
Let $X_1(t), \ldots, X_n(t)$ be functional samples observed over varying subsets, $\rS_1, \ldots, \rS_n$, of a compact set $\mathrm{C}$. Similar to \citet*{kraus2015, kraus2019} and \citet*{park2017}, we consider the observed curves to be the result of filtering latent full information curves $Y_1(t), \ldots, Y_n(t)$ on $\rC$ by independent indicator processes $\delta_1(t), \ldots, \delta_n(t)$, where
\[
\delta_i(t) =
\begin{cases}
1,& \text{if } Y_i(t) \text{ is observed;}\\
0,& \text{if } Y_i(t) \text{ is unobserved;}\\
\end{cases}
\]
for $t \in \rC$ and $i=1,\ldots,n$. We make general assumptions about the nature of the filtering functions $\delta_i$ and the modeling assumptions include the following:
\begin{itemize}
\item[M1:] The stochastic processes, $(Y_i,\delta_i):=\{(Y_i(t),\delta_i(t))$, $t\in \rC\}$, $i=1,\ldots,n$ are independent and identically distributed on $(\Omega, \mathscr{F},\mathbb{P})$ and jointly $\mathscr{F}$-measurable.
\item[M2:] There are missing sampling variables $V_{i} = (V_{i1},\dots,V_{iK}) \in \cV$ and there is a measurable missing scheme $h : \rC \times \cV \to \{0, 1\}$ such that: (i) $V_{1},\dots,V_{n}$ are i.i.d. random variables with common distribution $f$; (ii) $\delta_{i}(t) = h(t, V_{i})$.

\item[M3:] $E(\delta_{i}(t))=b(t)$, $t \in \rC$, where $b(\cdot)$ is
uniformly continuous and bounded away from zero, $\inf_{t \in \rC} b(t) > 0$.
\item[M4:] $Y_i(t)$ and $\delta_i(t)$ are independent for $i=1,\ldots,n$.
\end{itemize}

An advantage of using robust estimators here is to avoid restrictive moment assumptions on the process $Y_i$, enabling analysis of partially observed processes from heavy-tailed or outlier-prone sampling distributions. Condition M2 is used for proving the uniform convergence of the average of sample indicator processes $\delta_{i}(t), i =1,\dots,n$, to $b(t)$. \citet*{kraus2019} specified such sup-norm convergence of the averaged sample indicator processes as one of conditions. Here we only impose mild explicit conditions to derive large sample properties of the robust estimator, see Section \ref{sec:Theory}. This condition is satisfied by a wide range of partial sampling structures including the examples below.

\begin{ex}[Functional segments over random subintervals in $\rC$] \label{missingex1}
Define a random interval $S_i=[l_i, u_i] \subset  \rC$, where $l_i=\text{min}(v_{i1},v_{i2})$, $u_i=\text{max}(v_{i1},v_{i2})$, and the $v_{ij}$, $j=1,2$, are i.i.d. replicates of a random variable $V$ supported on $\rC$. Then Condition M2 is satisfied with $h(t, v_{i}) = \vone(l_{i} \leq t \leq u_{i})$. This framework can be extended to multiple random intervals per curve, $S_i= \cup_{k=1}^K [l_{ki}, u_{ki}]$, with $l_{ki}$ and $u_{ki} $, $k=1,\ldots,K,$ are i.i.d. from $V$. The latter corresponds to an example of fragmented functional data considered in \citet*{delaigle2020}.
\end{ex}

\begin{ex}[Functional segments over fixed subintervals in $\rC$]\label{missingex2}
Given a fixed set of intervals, $I_1,\ldots,I_m$ such that $\cup_{j=1}^m I_j=\rC$, we can define $h(t, v_i)=\vone ( t \in I_{v_i})$ with $v_i$ i.i.d. from a uniform discrete random variable $V$ on $\{1,\ldots,m \}$. The resulting scheme comprises a set of functional fragments observed over pre-determined subintervals as in the motivating example.
\end{ex}

\begin{ex}[Dense functional snippets (\citet*{lin2020a})]\label{missingex3}
Define an interval $S_i=[l_i, l_i+d] \subset \rC=[0,1]$, where $0 < d < 1$ denotes the length of subinterval, and the $l_i$ are  i.i.d. copies of a random variable $V_1$ with the support $[0, 1-d]$. Then Condition M2 holds with $h(t, v_{i}) = \vone ( l_i \leq t \leq l_i + d )$. A further extension is to allow $d_i$, the subinterval length, to be drawn from a distribution supported on [0,1], and then let $h(t, \mvv_{i}) = \vone ( l_i \leq t \leq l_i + d_i )$.
\end{ex}

Condition M3 implies that the full range is covered by a sufficient portion of the data for sufficiently large sample sizes. For example, in the case of random interval  $S_i=[l_i, u_i]$, the support of $V$ should have positive probabilities at both boundaries of $\rC$ to ensure positive $b(t)$ bounded away from zero. Lastly, letting $P$ denote the joint probability measure for $(Y, \delta)$, Condition M4 implies that $P=P_Y \cdot P_\delta$, where $P_Y$ and $P_\delta$ denote the marginal probability measures for $Y$ and $\delta$ on $\rC$, respectively. Along with Condition M3, it enables the estimation of the functional parameter of $Y$ based on the partially observed functions $X$.

\subsection{Marginal M-estimator  }
\label{sec:MMest}
For partially observed samples $X_1(t),\ldots, X_n(t)$, we define the functional M-estimator $\hat\theta_n(t)$ under the cross-sectional approach by minimizing the criterion marginally for all values of $t$ in parallel as below,
\begin{equation}\label{eq:thetahat}
\hat\theta_n(t)= \argmin_{h \in \mathbb{R}} \sum_{i=1}^n \delta_i(t) \rho\left( X_i(t) - h\right),
\end{equation}
for $t \in \rC$ satisfying $\sum_{i=1}^{n}\delta_i(t)>0$, where $\rho(\cdot)$ is a real-valued loss function. Otherwise, the estimator is undefined. In other words, for fixed $t$, $\hat\theta_n(t)$ represents a pointwise M-estimator calculated based on the information observed at spatial location $t$. If we observe undefined $\hat\theta_n(t)$ at certain range in $\rC$ under finite sample size, it can be estimated through interpolation or smoothing methods when smoothness and continuity of $\hat\theta_n(t)$ is assumed. In practice, discretized partial curves are observed on fine grids and interpolation can be applied for the estimation.

\begin{ex}[Observation Weighted Mean Functions]
In the special case with $\rho(x)=x^2$ the estimator $\hat\theta_n(t)$ reduces to the weighted sample mean function,
$$
\bar{X}_\delta(t) = \frac{\sum_{i=1}^n \delta_i(t) X_i(t)}{\sum_{i=1}^n \delta_i(t)}, \quad t\in \rC.
$$
\citet*{kraus2019} showed the consistency and asymptotic Gaussianity of $\bar{X}_\delta$ for estimating the mean function of $Y$ if $Y$ and $\delta$ are independent, and assuming moment conditions on the underlying distribution of $X_i(\cdot)$.
\end{ex}

\begin{ex}[Response function quantiles]
\label{fquantile}
The spatially dependent median and quantile functions for the response function nonparametrically characterize the functional response distribution. The M-estimator framework enables consistent estimation of the quantile functions, $Q_\tau(\cdot)$, via sample estimates of the form,
$$
\hat{Q}_\tau(t) = \argmin_{h \in \mathbb{R}} \sum_{i=1}^n \delta_i(t) \rho_\tau\left( X_i(t) - h\right),
$$
where for the quantile function of order $\tau\in(0,1)$ we define
$
\rho_\tau(x) = x(\tau - I(x < 0)),
$
extending the univariate quantile estimator described, e.g., in \citet*{koenker2005}.
\end{ex}

For the general class of marginal M-estimators, the following conditions are employed for the loss function $\rho$ to ensure robustness of the estimator and to allow weaker distributional assumptions.

\begin{itemize}
\item[A1] $\rho: \mathbb{R} \rightarrow \mathbb{R}$ is continuous even function and strictly increasing on $\mathbb{R}^+$ with $\rho(0)=0$.
\item[A2] $\rho$ is at most linearly increasing in the tails, therefore $|\rho(x_1)-\rho(x_2)| \leq L|x_1-x_2|$ for some constant $L > 0$.
\item[A3] $\rho$ is differentiable and its derivative $\psi=\dot{\rho}$ is continuous.
\item[A4] The second order derivative $\dot{\psi}=\ddot{\rho}$ is almost everywhere differentiable and Lipschitz continuous, i.e., $|\dot{\psi}(x_1)-\dot{\psi}(x_2)|\leq K |x_1-x_2|$ for some constant $K > 0$.
\end{itemize}
To prove consistency we require only conditions A1 and A2. Conditions A2 and A3 imply that $\psi$ is bounded, which enables efficient estimation for heavy-tailed functional data without assuming moment conditions. Weak convergence is established assuming the additional conditions A3 and A4. More general conditions than A3 and A4 are needed, e.g., to establish the weak convergence of the functional quantile estimators, although a smoothed approximate quantile M-estimator is covered by these conditions.

When the loss function $\rho$ is differentiable, functional M-estimator marginally solves the estimating equation,
\begin{equation}\label{EE_n}
  \frac{1}{n} \sum_{i=1}^n \delta_i(t) \psi ( X_i(t) - \hat\theta_n(t))=0, \quad t \in \rC.
\end{equation}

The advantage of defining the M-estimator marginally is that it can use all available data to estimate the underlying functional location curve in cases where the data are missing in a piecewise fashion. This is in contrast to the functional M-estimator of \citet*{sinova2018}, defined for complete data as
$$
\hat\theta^{\bH}_n(\cdot) = \argmin_{h\in\bH} \frac{1}{n} \sum_{i=1}^n
\rho\left(\Vert Y_i(\cdot) - h(\cdot)\Vert_\bH\right),
$$
where $\Vert\cdot\Vert_\bH$ is a norm for Hilbert space $\bH$. The estimator $\hat\theta^\bH_n$ is not directly applicable to the partially observed functional data because $Y_i(\cdot)$ is not fully observed. As an alternative approach, following a referee's suggestion, $\hat\theta^\bH_n$ can be calculated based on reconstructed complete curves by adopting curve reconstruction method (\citet*{kneip2019}). We shall examine its estimation performance in simulation studies (Section \ref{sec:sims1}) with the comparison of results from the marginal M-estimator.

The marginal M-estimator developed here is applicable to partially observed functional data without intermediate steps and provides consistent estimates of functional location parameter under regularity conditions as we will see in Section \ref{sec:Theory}. Furthermore, under the complete data framework with $\delta_i(t)=1$, $i=1,\ldots,n,$ $t \in \rC$, the marginal approach offers an alternative wherein the robustness or outlier resistance of the estimator is locally controlled along with the function rather than on the overall norm of the function. This is further demonstrated in the simulation studies of Section \ref{sec:sims}.

\subsection{Fisher consistency and invariance properties  }
\label{EstandInv}
In this section, we define the functional location parameter, a theoretical version of $\hat\theta_n(t)$, and investigate its properties. Given the joint probability measure $P$ for $(Y,\delta)$, functional location parameter $\theta(t)$ is defined as,
\begin{equation}\label{theta2}
\theta(t)= \argmin_{h \in \mathbb{R}} E_{P}\big[ \delta(t)\rho\big( Y(t)- h \big) \big], \quad t \in \rC.
\end{equation}
Under Conditions A1-A3, $\theta(t)$ also marginally satisfies
\begin{equation}\label{EE_M}
E_P[\delta(t) \psi \big( Y(t)- \theta(t) \big)]=0, \quad t \in \rC.
\end{equation}
It will be shown below that, under general conditions, $\hat\theta_n(\cdot)$ converges uniformly to $\theta(\cdot)$ as $n$ increases, and furthermore, that $n^{-1/2}(\hat\theta_n(\cdot) - \theta(\cdot))$ is asymptotically a Gaussian process.

In the special case where $Y$ has symmetric marginal distributions, the M-estimator estimates the same location as the mean and median functionals, assuming those exist. This generalizes the familiar Gaussian framework.
We assume $\Theta$ represents a functional parameter set in Riemann integrable $L^2(\rC)$ space, which includes piecewise continuous functions with a finite number of bounded jumps.
In particular, we have the following result.

\begin{prop}[Symmetric marginal distributions]
\label{symmetric}
Under conditions of M1, A1-A3, if the marginal density of $Y(t)$ for each $t \in \rC$ is symmetric about a deterministic function $\alpha(t) \in \Theta$, i.e., $Y(t)-\alpha(t)$ and $\alpha(t)-Y(t)$ have the same distribution, then $\theta(t)=\alpha(t)$.
\end{prop}
Proposition \ref{symmetric} implies that $\theta(t)$ represents the functional center when the marginal density of $Y(t)$ is symmetric for each $t \in \rC$. Next consider the special case where there is a shift location function such that subtracting the shift function from the function responses removes the spatial dependence of the marginal distribution of the response.
\begin{itemize}
\item[B1] [Shifted marginal stationarity] There exists a deterministic function $\alpha(t) \in \Theta$, such that the shifted process $Z(t)=Y(t)-\alpha(t)$ has constant marginal distributions, $Z(t) \sim F_Z$ for $t \in C$.
\end{itemize}

We then obtain the following proposition under generalized distribution of $Y$.
\begin{prop} [Shifted marginal stationarity]
\label{invariant}
Under conditions M1, M4, A1, A2, and B1 with the translation function $\alpha(t)$, there exists a constant $c$ depending only on $F_Z$, such that $\theta(t)=\alpha(t)+c$.
\end{prop}
Consequently, in this setting $\theta(t)$ is a well-defined location parameter that can inherit any smoothness or bounded jumps up to an additive constant depending on $\alpha(t)$.

\begin{rem}[$\tau$-quantile function]
In general, without assuming either symmetry or shifted marginal stationarity, if $\rho$ is defined as in Example \ref{fquantile}, then $\theta(t)$ represents a $\tau$-quantile functional.
\end{rem}

\subsection{Robust functionals and influence functions  }
\label{sec:IF}
We define the weighted M functional,
\begin{equation}\label{Mfun}
M(t,h,P)= E_{P}\big[ \delta(t) \big\{ \rho\big( Y(t)- h(t) \big) -\rho\big( Y(t) \big)\big\} \big],
\end{equation}
and $\theta(t)$ equivalently marginally minimizes $M(t,h,P)$ for $t \in \rC$; cf. Section 3.2 of \citet*{huber2005} for the univariate case. Under Conditions A1-A2, the expectation in (\ref{Mfun}) exists for every probability measure $P$ and we assume the following general conditions:
\begin{itemize}
\item[D1] $\sup_{\theta\in \Theta}\sup_{t \in \rC} \vert M(t,\theta,{P}_n)-M(t,\theta,P)\vert \overset{p}{\to} 0$, where $P_n$ denotes a sequence of measures converges weakly to a measure $P$.
\item[D2] For every $\epsilon > 0$, $\inf_{\theta^* \in \Theta} \inf_{t\in\rC}  \{ M(t,\theta^*,P) - M(t,\theta,P): \vert\theta^*(t)-\theta(t)\vert > \epsilon \} > 0$.
\end{itemize}
Condition D1 requires the uniform convergence of weighted M-functional over the parameter space $\theta \in \Theta$ and $t \in \rC$. As an example, if $P_n$ denotes the empirical measure of $\{Y_i(t), \delta_i(t), t \in \rC \}_{i=1}^n$, then for given $\theta \in \Theta$, uniform convergence over $t \in \rC$ holds under Condition A2. In the univariate case, Chapter 5 of \citet*{vandervaart2007} provides other possible assumptions to replace uniform convergence over parameter space. Condition D2 implies that, for $t \in \rC$, only $\theta(t)$ yields a minimum value of $M(t,h,P)$, thus it is a well-separated point of minimum at $t$. It holds under Condition A1 and the derivation of Influence function and the large sample properties will be based on above conditions on functional M.

Outlier sensitivity of an estimator is often measured by the influence function (\citet*{hampel1974}). Using that technique here, we consider contaminated curve distributions that may show atypical behavior in two ways: extreme or outlier fluctuations in the process $Y$, or outlying behavior in the missing process, such as dependence between $Y$ and $\delta$. For convenience, let $T(P)(t)$ denote the distribution-dependent functional corresponding to the parameter $\theta(t)$ .
We then consider the behavior of $T$ for contaminated distributions of the form
\begin{equation}\label{contamP}
P_\varepsilon = (1- \varepsilon) P + \varepsilon \Delta_{(Y^*,\delta^*)}
\end{equation}
where $\Delta_{(Y^*, \delta^*)}$ is a point mass distribution concentrated on $(Y^*, \delta^*)$.

We first establish the continuity of $T$ uniformly over the contaminating distribution, a robustness property that holds when the score function $\psi$ is bounded. Note that, by definition, $P_\varepsilon$ converges weakly to $P$ as $\varepsilon \to 0$.

\begin{thm} [Contamination Robustness]
\label{uniform_conti}
Conditions M1, M4, A1-A2, D1-D2 imply
\[
\lim_{\varepsilon \downarrow 0} \sup_{t\in C, (Y^*, \delta^*)}
\vert T(P_\varepsilon)(t) - T(P)\vert = 0.
\]
\end{thm}

Next we extend the notion of functional influence function, adapting the definition of \citet*{gervini2008} as, $IF_T(Y^*, \delta^*)(t) = \lim_{\varepsilon \downarrow 0} \varepsilon^{-1}\{ T(P_{\varepsilon})(t)- T(P)(t) \}$, if the limit exists, where $P_{\varepsilon}$ is given in (\ref{contamP}). 
The corresponding gross-error sensitivity (cf. \citet*{gervini2008}) with the sup-norm metric is given by, $\gamma_T^\infty=\sup\{ \sup_{t \in \rC}|IF_T(Y^*, \delta^*)(t)| : \mbox{any}~(Y^*, \delta^*) \}$.  
\begin{thm}[Influence Robustness]
\label{IF}
Under M1, M4, A1-A4, if we assume uniform continuity of the functional $T(P)(t)$ and $\inf_{t\in\rC}$ $|E_P \big[ \delta(t) \dot\psi(X(t),\theta(t))\big] |>0$, then
\begin{equation}
IF_T(Y^*, \delta^*)(t)=\frac{\delta^*(t)\psi(Y^*(t),\theta(t))}{-E_P \big[ \delta(t)\dot\psi(X(t),\theta(t)) \big]}, \quad t \in \rC,
\end{equation}
and the boundedness of $\psi$ implies $\gamma_T^\infty < \infty$.
\end{thm}
Hence, boundedness of the marginal score function $\psi$ implies the bounded effect of heavy-tailed variation or outliers in the process $Y$ or the dependent missing process on the functional location parameter.

\section{Large Sample Approximations}
\label{sec:Theory}
\subsection{Uniform Consistency }
\label{sec:Cont}
In establishing consistency and asymptotic Gaussian approximations for the class of functional M-estimators, a key step is to develop an entropy bound used to establish sup-norm convergence for the averaged indicator processes $\delta(t)$. In particular, we establish the convergence of
\begin{equation}
\label{Wn}
W_n = \sup_{t \in \rC} \left\vert n^{-1} \sum_{i=1}^n [\delta_i(t) - b(t)] \right\vert,
\end{equation}
where, marginally for each $t \in \rC$, $\delta_i(t)\sim \Ber(b(t))$ are i.i.d., and the functions $t \mapsto \delta_i(t)$ are sampled from a general class on $\rC$ satisfying Condition M2.

To bound the size of $W_{n}$, we need to control the size of the function class, $\cG := \{h(t, \cdot) : t \in \rC \}$. Under the missing data sampling scheme in Condition M2, given a missing scheme $h : \rC \times \cV \to \{0, 1\}$, for any $g \in \cG$, there is a $t \in \rC$ such that $g(v) = h(t, v), v \in \cV$. Let $H : \cV \to \{0, 1\}$ be a measurable envelope for $\cG$, i.e., $H(v) \geq \sup_{g \in \cG} g(v) = \sup_{t \in \rC} h(t, v)$ for all $v \in \cV$. Define the uniform entropy integral as in \citet*{vandervaartwellner1996},
\begin{equation}
\label{eqn:uniform_entropy_integral}
J(\delta, \cG, H) = \int_{0}^{\delta} \sup_{Q} \sqrt{1 + \log(N(\cG, L^{2}(Q), \varepsilon \|H\|_{Q,2}))} \, \rd \varepsilon,
\end{equation}
where the supremum runs over all finitely discrete probability measures on $(\cV, \cB(\cV))$ and $N(\cG, L^{2}(Q), \varepsilon)$ is the $\varepsilon$-covering number of $\cG$ under the metric induced by $L^{2}(Q)$.

\begin{lem}[Expectation bound on $W_{n}$]
\label{lem:general_entropy_bound}
If Condition M2 holds and $J(1, \cG, H) < \infty$, then there is a universal constant $C > 0$ such that
\[
E[W_{n}]  \leq C {J(1, \cG, H) \over \sqrt{n}} \max\Big\{ 1, {J(1, \cG, H) \over \sqrt{n}} \Big\}.
\]
\end{lem}

\begin{cor}
\label{cor:entropy_bound}
(i) If $\cG$ is a finite class of functions (i.e., $|\cG| < \infty$), then the Hoeffding inequality and union bound imply that $J(1,\cG,H) \lesssim \sqrt{\log|\cG|}$, and Lemma~\ref{lem:general_entropy_bound} yields
\[
E[W_{n}] \lesssim \sqrt{\log{|\cG|} \over n} \max\Big\{ 1, \sqrt{\log{|\cG|} \over n} \Big\} \lesssim {1 \over \sqrt{n}}.
\]
(ii) If $\cG$ is a VC type class, i.e., there are constants $A, v > 0$ such that
\[
\sup_{Q} N(\cG, L^{2}(Q), \varepsilon \|H\|_{Q,2}) \leq \Big( {A \over \varepsilon} \Big)^{v} \quad \mbox{for all } \varepsilon \in (0, 1],
\]
then
\[
J(\delta, \cG, H) \lesssim \delta \sqrt{v \log\Big({A \over \delta}\Big)} \quad \mbox{for all } \delta \in (0, 1].
\]
Then Lemma~\ref{lem:general_entropy_bound} implies that there is a constant $C(v,A) > 0$ depending only on $v$ and $A$ such that
\[
E[W_{n}] \leq {C(v, A) \over \sqrt{n}}.
\]
In either cases (i) or (ii), we get the uniform rate of convergence $n^{-1/2}$ for estimating $b(t)$ by $n^{-1} \sum_{i=1}^{n} \delta_{i}(t)$, i.e.,
\[
E [W_n] = O(n^{-1/2}).
\]
\end{cor}

\begin{rem}
Example \ref{missingex2} is corresponding to the case (i), a finite class of functions $\cG$, and Example \ref{missingex1} and Example \ref{missingex3} are examples of case (ii),  a VC type class of $\cG$, thus $E[W_n]=O(n^{-1/2})$ holds for all the examples presented above.
\end{rem}

Based on Lemma~\ref{lem:general_entropy_bound}, the following result establishes uniform consistency of the M-estimator.
\begin{thm}[Uniform consistency]
\label{thm:uniform_conv}
 Under conditions of M1-M4, A1, A2, D1-D2, $\hat \theta_n(t)$ converges to $\theta(t)$ uniformly over $t \in \rC$.
\end{thm}

\begin{rem}
As a special case, we obtain the uniform consistency of the functional quantile estimators of Example 5 for partially observed functional data.
\end{rem}

\subsection{Functional Central Limit Theorem }
\label{sec:FCLT}
We first derive a general functional central limit theorem for functional sample mean under the missing data framework, previously studied by \citet*{park2017} and \citet*{kraus2019}, then adapt the result to obtain an asymptotic Gaussian process approximation for the proposed M-estimators. Let $\rC$ be a compact subset in a general metric space equipped with a metric $d$ and $V(t)$, $t \in \rC$, be a mean-square continuous process defined on a probability space $(\Omega, \sF, P)$. We suppose that $V(t,\cdot)$ is measurable for each $t \in \rC$, and $V(\cdot,\omega)$ is continuous for each $\omega \in \Omega$. We consider the second-order stationary process $V$ with mean zero and the covariance function $\gamma$ (i.e., $\gamma(s,t) = \Cov(V(s), V(t)), s,t \in \rC$), denoted by $V \sim \SP(0,\gamma)$. We define the process $Z_n(t)$ as,
\[
Z_{n}(t) = {\sqrt{n} \sum_{i=1}^{n} \delta_{i}(t) V_{i}(t) \over \sum_{j=1}^{n} \delta_{j}(t)}, \quad t \in \rC.
\]
The following result is adapted from a functional central limit theorem of \citet*{kraus2019}, which specified a key step, sup-norm convergence of the averaged sample indicator processes in \eqref{Wn}, as one of technical conditions. But here we establish asymptotic Gaussianity through Lemma \ref{lem:general_entropy_bound} under more explicit and practical Condition M2.

\begin{thm}[Functional Central Limit Theorem for partially observed data]
\nonumber
\label{thm:fCLT_missing_data_generic}
Let $V_{1},\dots, V_{n}$ be i.i.d. samples of the second-order stationary process $V$. Under M2-M4 with replacement of $Y$ by $V$, we have
\begin{equation}
\label{eqn:fCLT_missing_data_generic}
\nonumber
Z_{n} \rightsquigarrow \GP(0, \vartheta),
\end{equation}
where $\vartheta(s,t) = \gamma(s,t) v(s,t)b(s)^{-1} b(t)^{-1} ,~ s,t \in \rC $ and $v(s,t) = E_{P_\delta}[\delta(s) \delta(t)]$.
\end{thm}

\subsection{Gaussian Process Approximation of M-Estimator }
\label{sec:MCLT}
Building on the uniform consistency of the marginal M-estimators and the functional central limit theorem, the results of this section establish that robust M-estimators have Gaussian process large sample approximations under weaker distributional conditions than the functional sample mean. For notational simplicity, we denote $\psi(x-\theta)$ by $\psi(x,\theta)$.

\begin{thm}[Asymptotic normality of M-estimator]
\label{thm:MEST-AN}
Under conditions M1-M4, A1-A4, D1-D2, and if $E_{P_Y} [\dot{\psi}( Y(t),\theta(t))]$ exists and non-singular almost everywhere on $\rC$,
\[
\sqrt{n}\big( \hat \theta_n(t) - \theta(t) \big) \rightsquigarrow  GP (0, \xi),
\]
$\xi(s,t)=\varphi(s,t) E_{P_Y}[\dot{\psi}(Y(s),\theta(s))]^{-1} E_{P_Y}[\dot{\psi}(Y(t),\theta(t))]^{-1}$, where $\varphi(s,t)$ $=$ $\cov\big\{\psi(Y(t),\theta(t)),$ $ \psi(Y(s),\theta(s))\big\} v(s,t)b(s)^{-1} b(t)^{-1}$ with $v(s,t) = E_{P_\delta}[\delta(s) \delta(t)]$.
\end{thm}

\subsection{Robust Inferences}
\label{sec:TREND}
The uniform Gaussian approximation provides a tool for developing (i) robust functional ANOVA (fANOVA) test for equality of location parameters in several populations, and (ii) trend test to detect functional trends in observed curves. Specifically, we can follow-up the fANOVA type test by performing the trend test to see whether or not there is a specific systematic trend over $t$ in group differences, for example, constant or linear trend in the intergroup differences across the functional domain. 

\subsubsection{$L_2$ Test on Location Functions }
\label{sec:L2}
Statistical tests on robust location functions can be developed using the preceding asymptotic results. An important example is testing the equality of location functions in several populations with the null hypothesis $H_0: \theta_1(t)=\ldots=\theta_k(t)$, with $\theta_g(t)$, $g=1,\ldots,k$, representing functional location parameter \eqref{theta2} of population $g$. 

Under fully observed functional data structures \citet*{Shen2004}, \citet*{cuevas2004}, \citet*{Zhang2014} developed robust functional ANOVA tests of this type. Under partially observed data structures, \citet*{kraus2019} developed a functional mean-based fANOVA test on functional population mean. Our applications motivate robust testing for partially observed data, combining the two different issues investigated by the previous authors.

Let $X_{g1}(t), \ldots, X_{gn_g(t)}$, $g=1,\ldots,k$, denote partially observed functional curves for group $g$ and assume $\rC=[0,1]$ without loss of generality. Extending fANOVA under balanced data by \citet*{Shen2004}, we derive $L^2$-norm based test for testing equality of robust location functions with a test statistic, $T_n= \int_{t \in \rC} SSR_n(t)dt/ trace(\hat\xi)$,
where $n=\sum_{g=1}^k n_g$, $SSR_n(s)=\sum_{g=1}^k n_g[\hat\theta_{g}(t)- \bar \theta_{\cdot}(t)]^2$ with the functional M-estimator $\hat \theta_g$ for group $g$, and grand mean $\bar \theta_\cdot(t)= \sum_{g=1}^{k} n_g \hat\theta_g/n$. Here $\hat\xi$ represents consistent estimator of asymptotic covariance of the functional M-estimator in Theorem \ref{thm:MEST-AN}.
\begin{cor}
\label{cor:ANOVA_test}
Assume that $n_g \rightarrow \infty$, $n_g/n=a_g>0$ for $g=1,\ldots,k$, $trace(\xi)<\infty$, and $\xi(t,t)>0$, for any $t\in \rC$, where $\xi(s,t)$ denotes the asymptotic covariance function of the functional M-estimator derived in Theorem \ref{thm:MEST-AN}. Then under the null hypothesis of equal location functions and under the same conditions of Theorem \ref{thm:MEST-AN}, we have $T_n \convD T_0$, where
\[
T_0 \myeq  \sum_{r=1}^{\infty} \lambda_r A_r,\quad A_r \mysim \chi^2_{k-1},
\]
where $\lambda_r=\kappa_r /trace(\xi),~r=1,...,\infty$, are the scaled eigenvalues with $\kappa_r$, $r=1,\ldots,\infty$, the decreasing-ordered eigenvalues of covariance function $\xi(s,t).$
\end{cor}
In practice, we calculate the test statistic by plugging in the estimated covariance function from bootstrap procedure, avoiding the complications associated with estimation of the asymptotic covariance function and its eigenvalues; see Section \ref{sec:BOOT}.

\subsubsection{Functional Trend Test }
The Gaussian approximation for the M-estimator functionals enables a corresponding approximation for inference on trends or probes even if the data are only partially observed as described above.
\begin{cor}
\label{cor:linear_operator}
Under the same conditions of Theorem \ref{thm:MEST-AN}, let $c=\langle \theta(\cdot),\phi(\cdot) \rangle$, where $\phi(\cdot)$ is a fixed Riemann integrable $L^2$ function on $\rC$ and $\langle \cdot,\cdot\rangle$ represents inner product of fixed functions over $\rC$, $\langle f,g \rangle$ = $\int_{\rC} f(t)g(t)dt$. Define  $\zeta_n=\sqrt{n}\Big( \langle \hat\theta_n(\cdot),\phi(\cdot)\rangle - c \Big)$. Then under $tr(\xi)<\infty$,
\begin{equation}
\nonumber
\zeta_n  \rightsquigarrow N(0,\tau^2 ),
\end{equation}
where $\tau^2=\int\int_{\rC} \phi(s)\xi(s,t)\phi(t)dsdt$.
\end{cor}

In the context of balanced, complete functional data, \citet*{ramsay2005} called parameters like $\langle \theta(\cdot),\phi(\cdot) \rangle$ probes, generalizing the concept of a contrast, and discussed asymptotic confidence intervals. Here we derive asymptotic confidence intervals for the partially observed data framework, and these intervals can provide information whether or not a trend of interest is present. We also can perform it as a follow-up test of robust fANOVA test to detect systematically distinct behaviors among functional location parameters. Examples of application are presented in Sections \ref{sec:sims} and \ref{sec:QUS}.

\section{Bootstrap Inference}
\label{sec:BOOT}

This section provides a bootstrap approach to perform the trend test. The key is to jointly resample the $Y$ and $\delta$ processes simultaneously. Under the assumption of missing completely at random in Condition M4, it might seem ideal to generate partially observed pseudo samples by resampling $Y$ and $\delta$ over $\rC$ independently, but this is impossible in practice due to the missing information on unobserved segments of $X_i(t)$, $i=1,\ldots,n$. Instead, we generate pseudo-observations by resampling partially observed curves from the data. The resulting bootstrap recovers the missing at random dependence structure asymptotically.

Suppose that $\vY(t)=[Y_1(t),...,Y_n(t)]^T$ and $\mvdelta(t)=[\delta_1(t),...,\delta_n(t)]^T$ are the observed $Y$ and $\delta$ information. Let $\vU=[\vU_1,..., \vU_n]^T$ denote $(n \times n)$ matrix, where $\vU_i$ $\sim$ $\text{multinomial}$ $(1,$ $\text{rep}(1/n,n))$. Here $\vU_i$ tells which functional curve is chosen for the $i$th bootstrap sample. Then the bootstrapped functional vector is generated by $\vY^*(t)=[Y^*_1(t),...,Y^*_n(t)]^T=\vU \vY(t)$ and $\mvdelta^*(t)$ $=$ $[\delta^*_1(t),...,\delta^*_n(t)]^T$ $=$ $\vU \mvdelta(t)$. The joint resampling is the use of the same $\vU_i$ to generate $Y^*_i$ and $\delta^*_i $, which is corresponding $X_i^*$ eventually, where $\vX^*(t)=[X^*_1(t),...,X^*_n(t)]^T=\vU \vX(t)$. The bootstrap algorithm for the robust $L^2$-test is as follows.

\begin{breakablealgorithm}
\floatname{algorithm}{Algorithm 1}
\renewcommand{\thealgorithm}{}
\caption{1. Bootstrap approximation for testing the equality location parameters}
\label{protocol1}
\begin{algorithmic}[1]
\STATE Calculate $\hat \theta_g(t)$ from observed samples $\{X_{gi}(t)\}_{i=1}^{n_g}$, $g=1,\ldots,k$ and calculate $\bar \theta_{\cdot}(t)=\sum_{g=1}^{k}n_g \hat \theta_g(t)/n$
\FOR {g=1,\ldots,k}
\FOR {b=1,\ldots,B}
\STATE Sample pseudo partially observed curves $\{X^*_{gi}(t)\}_{i=1}^{n_g}$ with replacement from $\{X_{gi}(t)\}_{i=1}^{n_g}$
\STATE Calculate $\hat \theta_g^{*(b)}(t)$ from the pseudo observations $\{X^*_{gi}(t)\}_{i=1}^{n_g}$
\ENDFOR
\ENDFOR
\STATE Based on $\hat \theta_g^{*(b)}(t)$, $g=1,\ldots,k$, $b=1,\ldots,B$, for $t \in \rC=[0,1]$,\\
 calculate $\hat \xi^*_n(t,t)=\frac{n}{kB} \sum_{g=1}^{k}\sum_{b=1}^{B}(\hat \theta_g^{*(b)}(t)-\bar \theta_g^{*(\cdot)}(t))^2$, where $\bar\theta_g^{*(\cdot)}(t)= \frac{1}{B} \sum_{b=1}^{B} \hat \theta_g^{*(b)}(t)$
\STATE Calculate the test statistic $T_n^*$ by replacing $trace(\hat\xi)$ with $\int_\rC\hat\xi^*_n(t,t) dt$
\STATE For $t ,s \in \rC$, calculate $\hat\xi^*_n(t,s)=\frac{n}{kB} \sum_{g=1}^{k}\sum_{b=1}^{B}(\hat \theta_g^{*(b)}(t)-\bar \theta_g^{*(\cdot)}(t))(\hat \theta_g^{*(b)}(s)-\bar \theta_g^{*(\cdot)}(s))$ and calculate $\lambda_r^*$, $r=1,\ldots, R$, in Corollary \ref{cor:ANOVA_test} based on bootstrapped eigenvalues $\kappa_r^*$, $r=1,\ldots, R$ of $\hat\xi^*_n(t,s)$ and bootstrapped trace $\int_\rC\hat\xi^*_n(t,t) dt$
\STATE Approximate the $p$-value of location test based on $T_n$ and $T_0^*=\sum_{r=1}^{R} \lambda^*_r \chi^2_{k-1}$
\end{algorithmic}
\end{breakablealgorithm}

\begin{rem} In the procedure, we plug in the bootstrap variance $\hat\xi^*_n(t,t)$ to calculate $T_n^*$ and use this as a test statistic. Under the joint bootstrap, for $t \in \rC$, conditional second moments of bootstrapped samples $\Cov(\rho(Y_i^*(t))|\vY(t))$ and  $\Cov(\delta_i^*(t)|\mvdelta(t))$ converge to $\Cov(\rho(Y_i(t)))$ and $\Cov(\delta_i(t))$, respectively, as $B, n \rightarrow \infty$. Also, $Y_i^*(t)$ and $\delta_i^*(t)$ are asymptotically uncorrelated with correlation approaching to zero as $B, n \rightarrow \infty$. Thus, $T_n^*$ converges to $T_n$ by Slutsky's theorem. In the same manner, $T_0^*$ converges to $T_0$ and it justifies the use of bootstrapped procedure for robust inferential test. The simulations of Section \ref{sec:sims} provide empirical confirmation of the accuracy of the bootstrap inference.
\end{rem}

\begin{breakablealgorithm}
\floatname{algorithm}{Algorithm 2}
\renewcommand{\thealgorithm}{}
\caption{2. Bootstrap confidence interval for functional trend}
\begin{algorithmic}[1]
\FOR {b=1,\ldots,B}
\STATE Sample pseudo partially observed curves $\{X^*_i(t)\}_{i=1}^n$ with replacement from $\{X_i(t)\}_{i=1}^n$
\STATE Calculate M-estimator $\hat\theta^{*(b)}_n(t)$ from the pseudo observations $\{X^*_i(t)\}_{i=1}^n$
\STATE Project $\hat\theta^{*(b)}_n(t)$ to the direction of interest $\phi(t)$ and calculate $\hat c^{*(b)}$.
\ENDFOR
\STATE Based on $\hat c^{*(b)}$, $b=1,\ldots, B$, calculate $(1-\alpha)100\% $ bootstrap confidence interval via $100\alpha/2$-th and $100(1-\alpha/2)$-th percentiles of bootstrap distribution $\hat\theta^{*}_n(t)$
\end{algorithmic}
\end{breakablealgorithm}

If the bootstrapped confidence interval of the projection coefficient excludes zero, the test for trend is significant, otherwise it is not.

\section{Simulation Study}
\label{sec:sims}

We present two simulation studies. In Section \ref{sec:sims1}, finite sample behavior of the M-estimator is examined by comparing the estimation accuracy of our marginal approach to that of existing functional approaches (i) under complete functional data without missing and (ii) under partially observed functional data. Especially, for incomplete case, the functional estimators are applied to reconstructed curves. In Section \ref{sec:sims2}, we investigate the performance of the bootstrap inference for trend test with different structures of partially observed functional data.

\subsection{Simulation Study: Estimation Accuracy }
\label{sec:sims1}
We first generate functional data under six scenarios to investigate the estimation accuracy and perform comparative study. In total 80 independent curves following $X(t)=\mu(t)+ \sigma(t)\epsilon(t)$, $t \in [0,1]$, are generated for each scenario by varying assumptions on $\sigma(t)$ or $\epsilon(t)$, or by adding artificial contamination under fixed smooth location function $\mu(t)$. Here, $\sigma(t)$ represents magnitude of the noise and $\epsilon(t)$ denotes the error process. The goal is the estimation of $\mu(t)$ under various settings. Examples of simulated data under the different scenarios are shown in the Supplementary Materials.

For models 1-3, we generate data with the Gaussian, $t_3$, and Cauchy processes assumed on $\epsilon(t)$, respectively, with the constant $\sigma(t)=2$ over $[0,1]$. The exponential spatial correlation structure is assumed on noise process, where $\mbox{Cor}(\epsilon(t_1),\epsilon(t_2))=\exp(-|t_1-t_2|/d)$. Here, the range parameter $d$ determines the spatial dependence within a curve and the value of $0.3$ is used, but the studies with other values show the similar results. All curves are simulated at 100 equidistant points in $[0,1]$. 

Model 4 considers the data with $t_3$ white noise error with random scales, where $\sigma(t)$, $t \in [0,1]$, is generated from $N(2,10^2)$. Models 5 and 6 generate the partially contaminated data, where $X(t)=\mu(t)+\sigma(t)\epsilon(t)$, for $ t \in [0,0.2) \cup (0.4,1]$, under Gaussian process with the constant scale as in model 1-3, and $X(t)=\mu(t)+\zeta(t)$, for $t \in [0.2,0.4]$. For model 5, we consider Cauchy distributed white noise error process $\zeta(t)$ with unit scale and, for model 6, Cauchy distributed contamination under exponential spatial correlation is assumed with unit scale.

\begin{table}[b!]
\caption{Relative median ISE of the mean with respect to that of the proposed M-estimator under unscaled robust tuning parameter for models 1-6.}
\centering
\begin{tabular}{cccccc}
\hline
model1 & model2 & model3 & model4 & model5 & model6 \\
\hline
0.78 & 1.50 & 79.9 & 1.64 & 98.6 & 16.4 \\
\hline
\end{tabular}
\end{table}

To calculate the proposed M-estimator, we use the Huber loss function and estimate the location parameter using constant or scaled robust tuning parameters. The former one uses the constant tuning parameter, say $c$, on $t \in [0,1]$, and the latter one applies varying robust tuning parameters, $c(t)=r*\text{MAD}(t)$, $t \in [0,1]$, where $r$ controls the overall robustness and $\text{MAD}(t)$ indicates marginal median absolute deviation (MAD) of the response function values at $t$. In the simulation, we choose $c$ as 0.8 to make the marginal estimates as close as marginal median values. For scaled approach, we set $r=0.2$ to make a fair comparison considering that $\sigma=2$.

For the comparative study, two competitors, a functional M-estimator developed by \citet*{sinova2018} and a functional median proposed by \citet*{gervini2008} are considered. Under complete data without missing, they can be directly applied to generated data using the same Huber loss function with a robust tuning parameter as $0.8$ for functional M-estimator.

To evaluate the performance, the integrated square error (ISE) is calculated, $\mbox{ISE}(\hat\mu)= \sum_{t=1}^{100}[\hat\mu(s_t)-\mu(s_t)]^2/100$ , over 500 repetitions.

\subsubsection{Estimation Accuracy for Complete Functional Data }
\begin{figure}[t!]
  \centering
  \includegraphics[width=6in]{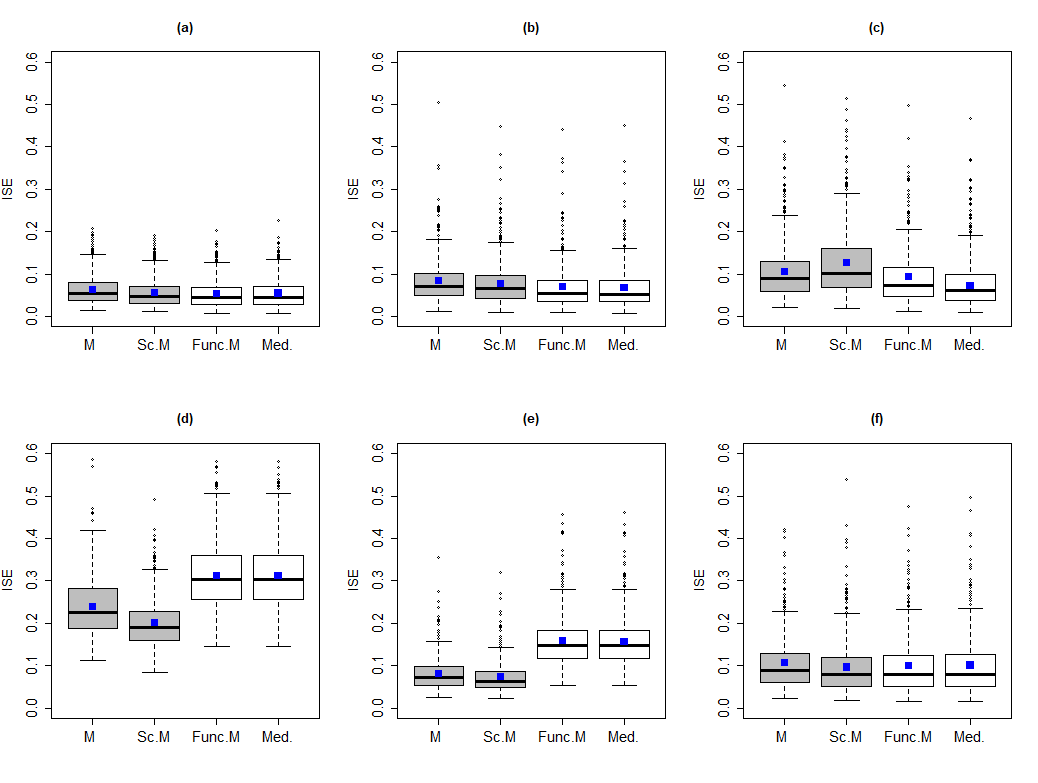}
  \caption{Boxplots of ISE over 500 replications from the marginal M-estimator (M), marginal scaled M-estimator (Sc.M), functional M-estimator (Func.M), and functional Median (Med.) under completely observed data from (a) Gaussian, (b) $t_3$, (c) Cauchy, (d) white-noise $t_3$ with random scales, Gaussian partially contaminated by (e) Cauchy white-noise, and by (f) Cauchy processes. Blue square dots represent mean values.}\label{simulation_IMSE}
\end{figure}

Estimation accuracies under complete data are displayed in Figure \ref{simulation_IMSE}. Here, two grey boxes represent results from the marginal M-estimators with `M' and `Sc.M' denoting M-estimator with constant tuning parameter and with the scaled tuning parameter. The `Func.M' and `Med.' indicate functional M-estimator from \citet*{sinova2018}, and median from \citet*{gervini2008}. The results from the sample mean are excluded in visualization due to exceedingly large ISE's in all scenarios except Gaussian case. Instead, relative ratios of median ISE of the mean with respect to that of the proposed M-estimator are presented in Table 1. Similar relative ratios are found with respect to other robust estimators, thus they are not included in the paper.

\begin{figure}[t!]
  \centering
  \includegraphics[width=6in]{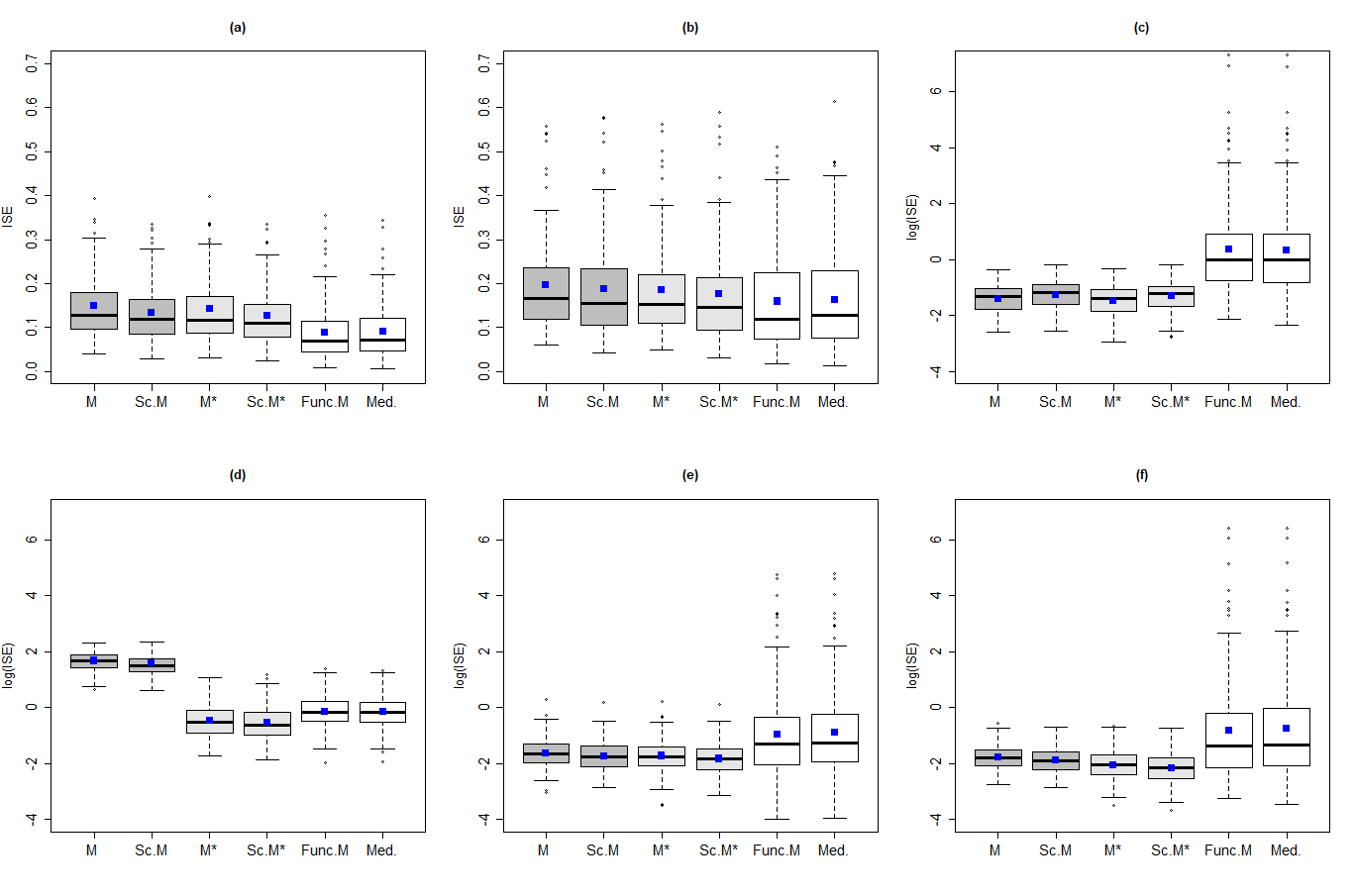}
  \caption{Boxplots of ISE or log transformed ISE over 500 replications from the marginal M-estimator (M), marginal scaled M-estimator (Sc.M), marginal M-estimator under pre-smoothed curves (M*), marginal scaled M-estimator under pre-smoothed curves (Sc.M*), functional M-estimator (Func.M), and functional Median (Med.) under partially observed data over random intervals from (a) Gaussian, (b) $t_3$, (c) Cauchy, (d) white-noise $t_3$ with random scales, Gaussian partially contaminated by (e) Cauchy white-noise, and by (f) Cauchy processes. Blue square dots represent mean values.}\label{simulation_IMSE_randommissing}
\end{figure}

Under the Gaussian model (model 1), robust estimators achieve a similar estimation accuracy as functional mean does, but under the heavy-tailed or contaminated scenario, we observe the failure of the sample mean with large ratios. Now, for the comparison among robust estimators in Figure \ref{simulation_IMSE}, we see that under (a) Gaussian and (b) $t_3$ errors, all four estimators achieve the almost similar estimation accuracy. Under (c) Cauchy noise, existing functional approaches slightly outperform, but it is not surprising because the discretized curves are generated under multivariate Cauchy distribution where the discretized functional approach is meant to be optimal. But the marginal approach still achieves comparable performance. Under the constant $\sigma$ in model 1-3, M-estimator with constant tuning parameter seems slightly more stable than the estimator with a scaled parameter, but it does not seem to be a significant difference. The plot (d) displays the estimation error from the marginally independent noise and we see that two competitors fall behind the marginal approach in estimation accuracy. We observe the same pattern under the model of partial contamination by marginally independent Cauchy noise in (e). Under the spatially correlated contamination in (f), we again see a similar performance among four methods. Contrary to the comparable estimation errors among estimators under model 1-3, the distinction in performance is apparent under models 4 and 5. And, for the model of random noise scale, the M-estimator with scaled tuning parameter slightly outperforms one with unscaled parameter. In summary, our proposed marginal approach provides comparable or superior performances in estimation accuracy under various scenarios, compared to existing methods.

\subsubsection{Estimation Accuracy for Partially Observed Functional Data }
\label{sec:sims12}
To investigate the estimation performances under incomplete data with missing segments, we apply two sampling frameworks to each generated set of curves; (i) partial sampling process under random interval sampling (Example \ref{missingex1}), where, $v_{1i}$ and $v_{2i}$ being generated from Beta(0.3,0,3), and $\delta_i(t)=\vone( \min(v_{i1},v_{i2}) \leq t \leq \max(v_{i1},v_{i2}))$, $i=1,...,80$, and (ii) random missing process under fixed number of intervals (Example \ref{missingex2}), $I_j$, $j=1,2,3$, $t\in[0,1]$ satisfying $\cup_{j=1}^3 I_j=[0,1]$, and randomly assign one of three to $i$th curve, for $i=1,...,80$. Note that we perform the analysis hereafter for $t \in [\varepsilon, 1-\varepsilon]$ with $\varepsilon=0.01$ and scaled Beta distribution to ensure the Condition M3 as discussed in Section \ref{sec:model}.

As two functional competitors are not directly applicable to incomplete curves, we apply them to reconstructed curves by adopting consistent reconstruction technique from \citet*{kneip2019} as described in Section \ref{sec:MMest}. Note that \citet*{kneip2019} employ FPCA to reconstruct unobserved segments and consistent estimation is fulfilled by nonparametric smoothing, e.g., local linear smoother, on observed fragments. Thus, for a fair comparison, we apply the proposed M-estimators to pre-smoothed curves, especially under robust nonparametric smoothing (\citet*{Fried2007}). For each simulated set, the same bandwidth is used in both local linear smoothing and robust smoothing.

Figure \ref{simulation_IMSE_randommissing} presents estimation accuracy from partially sampled data under random intervals with two more estimators, `M*' and `Sc.M*', calculated based on pre-smoothed curves via robust smoothing. Results from partial sampling process under fixed number of intervals are provided in supplementary materials with very similar findings from random interval cases.

Boxplots from (a) Gaussian and (b) $t(3)$ show similar results that we observe from regular structured data with slightly lower errors from existing functional approach under Gaussian model. We also observe similar errors of marginal M-estimator under raw and robust pre-smoothed data. Next we find interesting results under other distributional settings and boxplots are generated in log transformed ISE due to some severely large ISE's calculated from competitors. Under (c) Cauchy distribution or Gaussian data with Cauchy contaminations, either (e) Cauchy fragmented or (f) Cauchy marginal noise, we examine the failure of Func.M and Med. based on reconstructed curves with very large ISE's ranges from 0.03 to 600. It is due to unstable reconstruction from heavy-tailed partial observations demonstrating the failure of regular smoothing technique along with FPCA for data with potential outliers. Under these settings, proposed M-estimator extensively outperforms with similar performance using smoothed or raw data. In contrast,  under (d) marginal $t(3)$ heterogeneous scales of noise, application of the functional approach for reconstructed curves works well through smoothing, as it alleviates marginal peaks with the use of nearby neighbors information. Nonetheless, the marginal M-estimator based on robust pre-smoothed data outperforms the functional approach based on ISE.

\subsection{Simulation Study: Robust Inference }\label{sec:sims2}
In the second simulation study, we investigate the validity of the bootstrap-based inference in functional trend test. The coverage probability and the length of the bootstrapped confidence intervals are investigated under five models and five sampling structures including partially observed framework. For the first three models, we borrow model 1-3 from the first simulation by assuming Gaussian, $t_3$, and Cauchy process, respectively, in error process, but with $\mu(t)=\phi_0(t)+2\phi_1(t)+0.5\phi_2(t)$, $t \in [0,1]$, where $\phi_0(t)$, $\phi_1(t)$, and $\phi_2(t)$ representing orthonormal constant, linear, and quadratic basis functions, respectively. The other two models follow the contamination scenario in the first simulation, Gaussian curves contaminated by Cauchy process on [0,0.3] and [0.7,1], respectively, with constant noise scale. For each scenario, 80 curves are generated at 100 equidistant points over $[0,1]$.

We consider two more partial sampling frameworks additional to two missing procedures considered in the previous simulation; (iii) dense functional snippet (Example \ref{missingex3}), with length of subintervals set as $d=0.2$ and $\delta_i(t) = \vone ( l_i \leq t \leq l_i + d )$, where $l_i$ being drawn from Unif(0, 0.8) following a part of simulation settings in  \citet*{lin2020a}, and (iv) fragmented functional data on sparse and irregular grid points. Although condition M2 requires careful extension, not covered in this article, to include sparse irregular sampling scheme, we examine the performance of robust inference for potential study of extension. The data generation process and results under sparse design are provided in supplementary materials.

\begin{figure}[t!]
  \centering
  \includegraphics[width=6in]{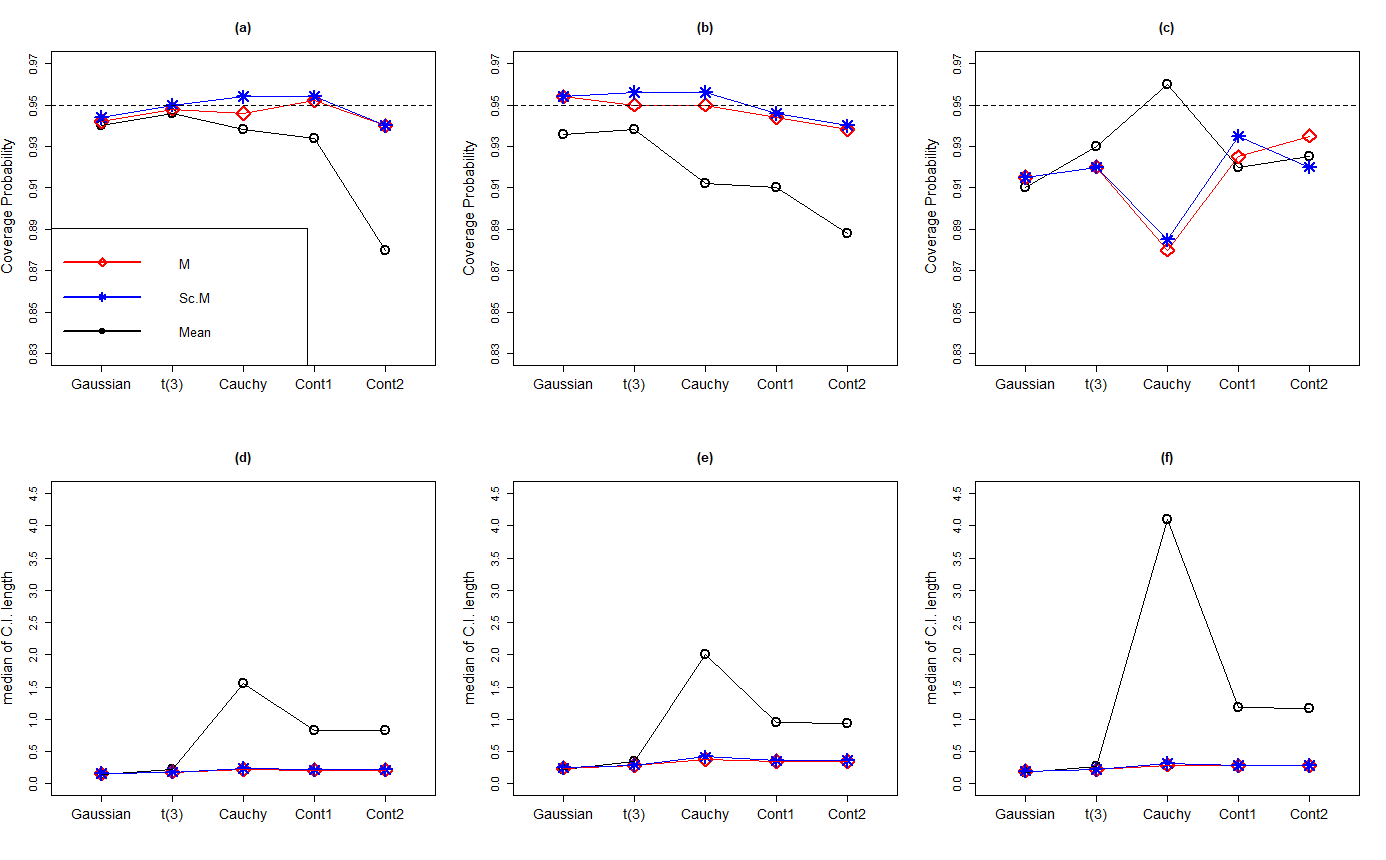}
  \caption{Coverage probabilities of bootstrapped confidence intervals of projection coefficients to quadratic function under Gaussian, $t_3$, Cauchy, and two contaminated data from M-estimator (M), scaled M-estimator (Sc.M), and Mean functions over 500 repetitions; (a) regular structure, (b) partially observed structure under random intervals, and (c) dense functional snippets. Median length of bootstrapped confidence intervals of projection coefficient under (d) regular structure, partially observed structure under (e) random intervals, and under (f) dense functional snippet }\label{fig:projection}
\end{figure}

We calculate bootstrapped $95\%$ confidence intervals of the projection coefficients to constant, linear, and quadratic functions under M-estimator, scaled M-estimator, and the sample mean function with 800 bootstrapped samples. Then the coverage probabilities are estimated from 500 repetitions based on the number of the inclusion of true coefficients in confidence intervals. Also, we calculate the median length of the intervals.

Figure \ref{fig:projection} displays results from the projected coefficients to the quadratic trend. Three plots in the first row illustrate the empirical coverage probabilities under regular, random interval, and dense functional snippet cases and results from other partial structures are provided in supplementary materials. In (a) and (b), we observe that the coverage probabilities from robust estimators are always around $95\%$, but the overall probabilities from mean tend to be less than the desired $95\%$. Especially, under the Cauchy or contaminated model, inference from the functional mean may fail to detect the true quadratic trend. In (c), we see coverage probabilities lower than desired $95\%$ but still around $90\%$ from our proposed inference and it is due to relatively small effective sample size at each $t$ with the length of subintervals set as $d=0.2$ in $[0,1]$. Interestingly, coverage probability, close to $95\%$ is observed through mean inference on Cauchy data, but it is no surprise considering the wide length of bootstrapped confidence intervals we shall see in (f). The plots of (d), (e), and (f) visualize the median length of confidence interval from each estimator, and the inference from functional mean seems less informative and unstable with the wide length of the interval. But the results from the proposed M-estimator seems stable regardless of distribution assumptions and missing structures.

\section{Example: Quantitative Ultrasound Analysis }
\label{sec:QUS}

We illustrate the estimation of M-estimator and inference with the Quantitative Ultrasound (QUS) data. As described in Introduction, \cite{wirtzfeld2015} presented data and results from diagnostic ultrasound studies using multiple transducers to scan mammary tumors (4T1) and benign fibrous masses (MAT) in rats and mice. In this experiment, total five transducers are used for noninvasive scan of each animal, and specifically, two transducers, 9L4 and 18L6, from Siemens, cover 3–10.8 MHz, L14-5 from Ultrasonix uses frequencies 3–8.5 MHz, and MS200 and MS400 from VisualSonics cover higher frequencies, 8.5–13.5 MHz, and 8.5–21.9 MHz. 
\begin{figure}[h!t!]
  \centering
  \includegraphics[width=5.5in]{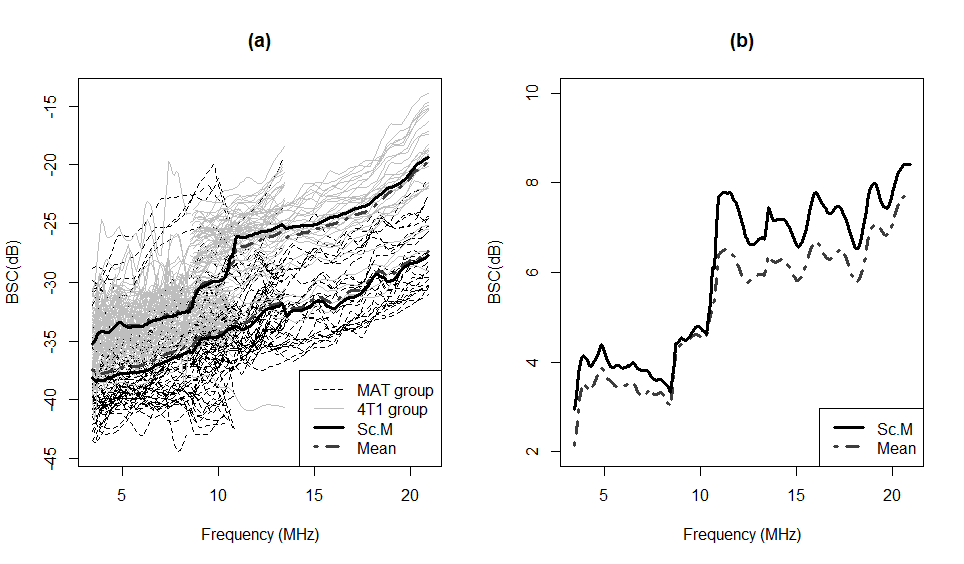}
  \caption{Quantitative Ultrasound data. (a) BSC curves from MAT and 4T1 tumors with proposed functional M-estimator and functional mean for each group. (b) Marginal group differences of M-estimator and mean.}\label{Tumor}
\end{figure}

The aims of this experiment are the detection of significant differences in the behavior of BSC curves between
two distinct tumors and investigation of the consistency among frequency ranges or transducers in such detection. To address this, we calculate the functional M-estimator and preform robust inferential tests.

Figure \ref{Tumor} (a) shows estimated group location parameters from marginal M-estimator under Huber loss with a scaled robust tuning parameter and from the functional mean for two tumor types. We observe remarkable jumps at 8.5 MHz and 10.8 MHz in a group of the 4T1 tumor and they are corresponding to frequencies where a change in the variety of transducers is observed. But the jump on functional mean at 10.8 MHz is weaker than the jump on M-estimator due to multiple outlying curves in the 4T1 group which have suspiciously small values or abnormal behaviors comparing to the majority.

To demonstrate significant distinction in behaviors of BSC from two different tumors, \cite{wirtzfeld2015} applied separate fANOVA to subsets of data, consisting of BSC curves collected from same transducer, spanning the same frequency ranges, to avoid partial sampling issue. Also they test the equality of functional mean parameters, which might not be valid with outlying curves. Thus we now perform $L^2$-type robust fANOVA test following bootstrap procedure in Section \ref{sec:BOOT} and significant group difference is detected with p-value $<.0001$ ($T_n^*=31.32$). It enables full scale analysis with higher power by using all curves in one test. Then, as a follow-up, we examine systematic trend in functional difference between two tumors and Figure \ref{Tumor} (b) presents seemingly increasing trend. But the inferential justification is needed to make a conclusion. At the same time, we also want to investigate the effect of transducers in BSC measures. To this end, we calculate bootstrapped confidence intervals of projection coefficients corresponding to the selected basis and step functions. We specifically consider constant, linear, and quadratic basis functions as well as three step functions, named as Step1, Step2, and Step3, where Step1 has a jump at 8.5 MHz, Step2 has a jump at 10.8 MHz, and Step3 at 13.5 MHz. Step functions are defined based on known transducer information. The inferences based on coefficients of the first three basis functions enable identifying a general trend, whether higher frequencies separate two groups more efficiently than lower frequencies do. Meanwhile, coefficients of three step functions provide information to discover the transducer effect. We adopt the Huber function in M-estimator with constant and scaled robust tuning parameters as discussed in Section \ref{sec:sims}. For unscaled one, we choose $c=0.8$, and for the scaled one, $r=0.4$ with the consideration of the overall estimated MAD over the whole frequency range.

\begin{table}[b!]
\myfontsizeA
\center
\caption{Estimated projection coefficients to basis functions. 95\% and 99\% bootstrapped confidence intervals in round brackets and square brackets, respectively. Bracket with * indicates an interval not including zero in it.}
\begin{tabular}{ccccccc}
  \hline
   & Quadratic & Linear & Constant & Step1 & Step2 & step3 \\
   \hline
   \multirow{ 3}{*}{\shortstack[l]{ \\M-estimator}}    & $-0.15$ & 1.55 & 6.00 & 0.22 & 0.66 &$ -0.11$\\
      & $(-0.52, 0.17)$ & $(0.54, 2.38 )^*$ & $(4.90, 6.97)^*$ & $( -0.05, 0.52)$ & $(0.29, 1.00)^*$ & $(-0.46, 0.31)$ \\
   & $[-0.65, 0.29]$ & $[0.18, 2.63]^*$ & $[4.51, 7.20 ]^*$ & $[-0.14,0.64 ]$ & $[0.18, 1.09 ]^*$ & $[-0.65, 0.44 ]$\\
   \hline

   \multirow{ 3}{*}{\shortstack[l]{~~Scaled \\ M-estimator}} & $-0.22$ & $1.48$ &$ 5.98$ & $0.24$ & 0.58 & $-0.06$\\
      & $(-0.50, 0.11)$ & $(0.56, 2.31)^*$ & $(4.88, 6.91)^*$ & $(-0.01, 0.49)$ & $(0.27, 0.86)^*$ & $(-0.42,0.26)$ \\
   & $[-0.61, 0.23]$ & $[0.20, 2.55]^*$ & $[4.50, 7.15]^*$ & $[-0.10, 0.56]$ & $[0.17, 0.97]^*$ & $[-0.55, 0.40]$\\
   \hline
   \multirow{ 3}{*}{Mean} & -0.16 & 1.32 & 5.28 & 0.24 & 0.34 & -0.07 \\
    & $(-0.44, 0.15)$ & $(0.58, 2.05)^*$  & $(4.33, 6.13)^*$ & $(0.02, 0.46)^*$ & $(0.05, 0.62)^*$ & $(-0.38, 0.25)$  \\
& $[-0.53, 0.24]$ & $[0.30, 2.29]^*$ & $[4.07, 6.35]^*$ & $[-0.05, 0.52]$ & $[-0.02, 0.71]$ & $[-0.47, 0.32] $\\
  \hline
\end{tabular}
\end{table}

Table 2 shows estimated coefficients of functional group difference projected to six basis functions and corresponding $95\%$ and $99\%$ bootstrapped confidence intervals based on 3000 replications. The discretized curves in the data are densely collected but do not share common grids, so interpolation is applied to each curve at an equally spaced grid of 176 points over 3-21.6 MHz. The computation time on 3.60GHz Intel(R) Core(TM) i7-7700 CPU is 234 seconds for derivation of bootstrapped confidence intervals from M-estimator with $n_{4T1}=115$, $n_{MAT}=149$.

First, we observe that results of M-estimator from scaled and constant tuning parameters look almost the same except the discrepancy in estimated coefficients of the quadratic term. But the quadratic trend is insignificant from both bootstrapped inferences, so fundamentally two estimators derive the same conclusion. Then a significant linear trend is detected in group differences with positive coefficients from M-estimator, which implies that higher frequencies are more efficient to detect group differences than lower frequencies are. The finding is the same for the mean approach, but shrunk estimate is observed due to the effect of outliers.

To examine the transducer effect, we see results from three step functions. Changepoint at 8.5 MHz (Step1) turns out to be insignificant from robust estimators, but 95\% confidence interval from mean does not include zero, implying significant distinct behavior at this jump. For the second change point, robust M-estimators detect significant positive jump at 10.8 MHz with confidence, with the lower bound far from zero, but the inference from mean function fails to detect such change in 99\% confidence interval. Although inference from 95\% confidence interval detects significant jump, it lacks confidence with lower bound very close to zero. Again, this different conclusion is due to multiple outliers in the 4T1 group and mean function underestimates the jump at this change point. The last change point between two VisualSonics transducers turns out to be insignificant from both estimators. In conclusion, BSC curves significantly distinct different tumors along with all frequency ranges and higher frequencies separate them more efficiently than lower frequencies do. Furthermore, we see a significant positive jump at 10.8 MHz, which implies the outperformed efficiency of VisualSonics transducers in terms of tissue distinction comparing to others.


\section{Discussion }
\label{sec:Conclusion }
We propose a class of robust M-estimator applicable to partially observed functional data. We show that our estimator is consistent and asymptotically follows the Gaussian process with root-$n$ rates of convergence under a key condition for sup-norm convergence of the indicator process. Also, robust inferential tools are developed under asymptotic normality and they can be performed in practice with bootstrap procedures. The validity of bootstrap inference is supported by convergence of conditional second moments of bootstrapped samples as well as simulation studies, where the true trend is detected with the desired coverage probability under heavy-tailed or contaminated distribution with various structures of missingness. In terms of estimation accuracy, numerical simulation experiments demonstrate how the proposed estimator can outperform existing functional robust estimators, even in the case of complete data.

The proposed partial sampling framework is particularly appealing as various types of recently emerged incomplete data structures satisfy assumptions our assumptions, including dense functional snippets (\citet*{lin2020a}) and fragmented functional data (\citet*{delaigle2020}). These connections demonstrate the wider applicability of the methods we developed here. In addition, based on our simulation studies, further extension to segmented data recorded at sparse and irregular design points is a promising direction for further development.

\section*{Supplementary Materials}
Online supplementary materials include (i) technical proofs of Propositions, Lemmas, and Theorems and (ii) figures and detailed results from simulation studies.
\par
\section*{Acknowledgements}
This work was supported by National Science Foundation CAREER Award DMS-1752614 (X. Chen), University of Illinois Research Board Award RB18099 (X. Chen) and National Institutes of Health Grant R01CA226528-01A1 (D. G. Simpson).



\par

\setcounter{section}{0}
\setcounter{equation}{0}
\def\theequation{S\arabic{section}.\arabic{equation}}
\def\thesection{S\arabic{section}}

\fontsize{12}{14pt plus.8pt minus .6pt}\selectfont

\section{Appendix: Proofs}
\label{proof}
\begin{proof} [\underline{Proof of Proposition \ref{symmetric}}]
Under M1, M4, A1-A3, we write the pdf of the marginal distribution of $Y$ at $t \in \rC$ as $f(y)$ and it is assumed to be symmetric (or even function) about $\alpha(t)$. Then
\[
E_{P_Y}[\psi\big(Y(t)-\alpha(t) \big)]=\int_{-\infty}^{\infty} \psi\big( Y(t)-\alpha(t) \big) f\big(Y(t)-\alpha(t) \big) dy=0,\quad t \in \rC,
\]
under the assumption of odd function $\psi(\cdot)$. Thus, $\theta(t)=\alpha(t)$.
\end{proof}

\begin{proof} [\underline{Proof of Proposition \ref{invariant}}]
Under A1-A2, equation \eqref{theta2} implies that $E_{P_Y}\big[\rho(Y(t)-\alpha(t) + \alpha(t) -\theta(t)) \big]$ equals specific value at each $t \in \rC$, say $c_1(t)$. Under B1, the marginal distribution of $Y(t)-\alpha(t)$, $t \in \rC$, does not depend on $t$ with the probability measure $P_Z$. Then we can equivalently write
\[
 E_{P_Z}\big[\rho(Z -\{ \alpha(t) -\theta(t)\}) \big]=c_1,
\]
and $\{ \alpha(t)-\theta(t) \}=c_1+c_2$, where constant $c_2$ is determined by $P_Z$. Let $c=c_1+c_2$ then we can write $\theta(t)=\alpha(t)+c$.
\end{proof}

\begin{proof}[\underline{Proof of Theorem \ref{uniform_conti}}]
Denote $T(P_\varepsilon)(t)$ by $\theta_\varepsilon(t)$. Under D2, for any $\upsilon>0$, there exists $\delta > 0$,
\begin{equation}
\nonumber
\begin{aligned}
P(\sup_{t\in\rC} |\theta_\varepsilon(t) &-\theta(t)| > \upsilon)\leq P(\sup_{t\in\rC} \big[ M(t,\theta_\varepsilon,P)-M(t,\theta,P)\big] > \delta )  \\
&\leq P(\sup_{t\in\rC} \big[ M(t,\theta_\varepsilon,P)-M(t,\theta_\varepsilon, P_{\varepsilon})+ M(t,\theta,P_{\varepsilon}) - M(t,\theta,P) \big] > \delta ) \\
  &\leq P(\sup_{t\in\rC} | M(t,\theta_\varepsilon,P)-M(t,\theta_\varepsilon, P_{\varepsilon})| > \delta/2 )\\
   & \qquad\qquad\qquad + P(\sup_{t\in\rC} |M(t,\theta,P_{\varepsilon}) - M(t,\theta,P)| > \delta/2 ).
\end{aligned}
\end{equation}
By D1, $T(P_{\varepsilon})(t)$ is uniformly continuous as $\varepsilon \to 0$.
\end{proof}

\begin{proof}[\underline{Proof of Theorem \ref{IF}}]
By the estimating equation of \eqref{EE_M},
\begin{equation}
\nonumber
\begin{aligned}
  0 &= (1-\varepsilon)E_P \big[ \delta(t) \psi(Y(t),\theta_\varepsilon(t))\big] + \varepsilon \delta^*(t) \psi(Y^*(t),\theta_\varepsilon(t)) \\
   &= (1-\varepsilon)E_P \big[ \delta(t)\{ \psi(Y(t),\theta_\varepsilon(t))-\psi(Y(t),\theta(t))\} \big]
   +\varepsilon \delta^*(t)\psi(Y^*(t),\theta_\varepsilon(t)) \\
   &= (1-\varepsilon)E_P \big[ \delta(t) \frac{\psi(Y(t),\theta_\varepsilon(t))-\psi(Y(t),\theta(t))}{\varepsilon}\big] + \delta^*
   (t)\psi(Y^*(t),\theta_\varepsilon(t))
\end{aligned}
\end{equation}
Let $\varepsilon \rightarrow$ 0, then
$$
0=E_P \big[ \delta(t) \dot\psi(Y(t),\theta(t))\big] \dot \theta(t)+ \delta^*(t)\psi(Y^*(t),\theta(t)).
$$
Thus,
$$
\dot \theta(t) = IF_T(Y^*,\delta^*)(t)= \frac{\delta^*(t)\psi(Y^*(t)-\theta(t))}{-E_P \big[ \delta(t) \dot\psi(Y(t),\theta(t))\big]},
$$
and the bounded $\psi(\cdot)$ implies $\gamma_T^\infty < \infty$.
\end{proof}


\begin{proof}[\underline{Proof of Lemma \ref{lem:general_entropy_bound}}]
Under the sampling scheme condition M2, we can define the empirical process
\[
\bG_{n}(t) = {1 \over \sqrt{n}} \sum_{i=1}^{n} [h(t, V_{i}) - \E h(t, V_{i})], \quad t \in \rC,
\]
where $V_{1},\dots,V_{n}$ are i.i.d. random variables in $\cV$ with common distribution $f$. Alternatively, we may write
\[
\bG_{n}(g) = {1 \over \sqrt{n}} \sum_{i=1}^{n} [g(V_{i}) - \E g(V_{i})], \quad g \in \cG
\]
with the identification of $g$ by $h_{t}$ for a given missing scheme $h$. Then
\[
W_{n} = {1 \over \sqrt{n}} \sup_{t \in \rC} \bG_{n}(t) = {1 \over \sqrt{n}} \sup_{g \in \cG} \bG_{n}(g).
\]
Recall that $H : \cV \to \{0, 1\}$ is a measurable envelope for $\cG$. Set $M = \max_{1 \leq i \leq n} H(V_{i})$. By the local maximal inequality \citet*{cck2014_empirical_process} with the locality parameter $\delta = 1$, there is a universal constant $C > 0$ such that
\[
E[\sup_{g \in \cG} \bG_{n}(g)] \leq C \Big\{ J(1, \cG, H) \|H\|_{f,2} + {\|M\|_{2} J^{2}(1, \cG, H) \over \sqrt{n}} \Big\}.
\]
Since $|H| \leq 1$ and $\|M\|_{2} \leq 1$, we get
\[
E[\sup_{g \in \cG} \bG_{n}(g)] \leq C \Big\{ J(1, \cG, H) + {J^{2}(1, \cG, H) \over \sqrt{n}} \Big\}.
\]
Then it is immediate that
\[
E[W_{n}] \leq C \Big\{ {J(1, \cG, H) \over \sqrt{n}} + {J^{2}(1, \cG, H) \over n} \Big\} \leq C {J(1, \cG, H) \over \sqrt{n}} \max\Big\{ 1, {J(1, \cG, H) \over \sqrt{n}} \Big\}.
\]
\end{proof}


\begin{proof}[\underline{Proof of Theorem \ref{thm:uniform_conv}}]
For $t \in \rC$, if $|\hat\theta_n(t)-\theta(t)|>\epsilon$, then $M(t,\hat\theta_n,{P})-M(t,\theta,P)>\delta_t$ by D2, and $\sup_t \big[ M(t,\hat\theta_n,{P})-M(t,\theta,P)\big] >\delta$, where $\delta=\sup_t \delta_t$. Then

\begin{equation}
\nonumber
\begin{aligned}
P(\sup_{t\in\rC} |\hat\theta_n(t)&-\theta(t)| > \epsilon) \leq P(\sup_{t\in\rC} \big[ M(t,\hat\theta_n,P)-M(t,\theta,P)\big] > \delta )  \\
 &= P(\sup_{t\in\rC} \big[ M(t,\hat\theta_n,P)-M(t,\hat\theta_n,P_n) + M(t,\hat\theta_n,P_n) - M(t,\theta,P_n) \\
 & \qquad\qquad\qquad + M(t,\theta,P_n) - M(t,\theta,P)\big] ) > \delta)\\
  &\leq P(\sup_{t\in\rC} \big[ M(t,\hat\theta_n,P)-M(t,\hat\theta_n,P_n)+ M(t,\theta,P_n) - M(t,\theta,P) \big] > \delta ) \\
  &\leq P(\sup_{t\in\rC} |M(t,\hat\theta_n,P)-M(t,\hat\theta_n,P_n)| > \delta/2 )\\
   & \qquad\qquad\qquad + p(\sup_{t\in\rC} |M(t,\theta,P_n) - M(t,\theta,P)| > \delta/2 )
\end{aligned}
\end{equation}
By D1, $\hat \theta_n(t)$ uniformly converges to $\theta(t)$ over $\rC$ as $n \rightarrow \infty$.
\end{proof}


\begin{proof}[\underline{Proof of Theorem \ref{thm:fCLT_missing_data_generic}}]
Let $\widetilde{Z}_{n}(t) = n^{-1/2} \sum_{i=1}^{n} \delta_{i}(t) V_{i}(t) / b(t)$. For any $t_{1},\dots,t_{K} \in \rC$, denote $\widetilde{\mvZ}_{n} = (\widetilde{Z}_{n}(t_{1}),\dots,\widetilde{Z}_{n}(t_{K}))^{T}$. By the multivariate CLT and the independence between $\delta_{i}$ and $V_{i}$, we have
\[
\widetilde{\mvZ}_{n} \stackrel{d}{\to} N(0, \Xi),
\]
where $\Xi = \{\vartheta_{jk}\}_{j,k=1}^{K}$ is the $K \times K$ covariance matrix with $\vartheta_{jk} = v(t_{j},t_{k}) \gamma(t_{j}, t_{k}) / [b(t_{j})b(t_{k})]$. By Theorem 7.4.2 in \citet*{HsingEubank2015}, the process $\{\widetilde{Z}_{n}(t) : t \in \rC\}$ is a random element in the Hilbert space $\bH = L^{2}(\rC, \cB(\rC), \mu)$, where $\mu$ is a finite measure on $\rC$. Then it follows from Theorem 7.7.6 in \citet*{HsingEubank2015} for i.i.d. Hilbert space valued random variables that
\[
\{\widetilde{Z}_{n}(t) : t \in \rC\} \rightsquigarrow \GP(0, \vartheta),
\]
where the finite-dimensional restrictions of $\vartheta$ is given by the covariance matrix $\Xi$. Note that
\[
\sup_{t \in \rC} \left| \widetilde{Z}_{n}(t) - Z_{n}(t) \right| \le \sup_{t \in \rC} |\widetilde{Z}_{n}(t)| \cdot \sup_{t \in \rC} \left| 1-{b(t) \over \overline{\delta}_{n}(t)}\right|,
\]
where $\overline{\delta}_{n}(t) = n^{-1} \sum_{i=1}^{n} \delta_{i}(t)$. Note that
\[
|\overline{\delta}_{n}(t)| \ge b(t) - |\overline{\delta}_{n}(t)-b(t)| \ge \inf_{t \in \rC} b(t) - W_{n},
\]
where
\[
W_{n} = \sup_{t \in \rC} |n^{-1} \sum_{i=1}^n [\delta_i(t) - b(t)]|.
\]
By Lemma 3.1, $E[W_n] =O(n^{-1/2})$.  Since $\sup_{t \in \rC} |\widetilde{Z}_{n}(t)| = O_{P}(1)$, we have
\[
\sup_{t \in \rC} \left| \widetilde{Z}_{n}(t) - Z_{n}(t) \right| = O_{P}(n^{-1/2}).
\]
Then Theorem \ref{thm:fCLT_missing_data_generic} is an immediate consequence of Slutsky's lemma.
\end{proof}

\begin{proof} [\underline{Proof of Theorem \ref{thm:MEST-AN}}]
The estimating equation \eqref{EE_M} can be equivalently written as,
\[
\frac{1}{\sum_{j=1}^n \delta_j(t)} \sum_{i=1}^{n}\delta_i(t) \psi(Y_i(t),\hat\theta_n(t))=0.
\]
By mean value theorem,
\[
\frac{1}{\sum_{j=1}^n \delta_j(t)} \sum_{i=1}^{n}\delta_i(t) \psi(Y_i(t),\theta(t)) +
\frac{1}{\sum_{j=1}^n \delta_j(t)} \sum_{i=1}^{n}\delta_i(t) \dot{\psi}(Y_i(t),\tilde\theta_n(t))\big( \hat\theta_n(t)-\theta(t)\big)=0,
\]
where $\theta(t) \leq \tilde \theta_n(t)\leq \hat\theta_n(t)$, $t \in \rC$. Rearranging terms, we get
\[
\sqrt{n}\big(\hat\theta_n(t)-\theta(t)\big)=- \Big[ \underbrace{\frac{1}{\sum_{j=1}^n \delta_j(t)} \sum_{i=1}^{n}\delta_i(t) \dot{\psi}(Y_i(t),\tilde\theta_n(t))}_{(1)} \Big]^{-1} \underbrace{\frac{1}{\sum_{j=1}^n \delta_j(t)} \sqrt{n}\sum_{i=1}^{n}\delta_i(t) \psi(Y_i(t),\theta(t))}_{(2)},
\]
where
\begin{equation}
\nonumber
\begin{aligned}
(1)&=\frac{1}{\sum_{j=1}^n \delta_j(t)} \sum_{i=1}^{n}\delta_i(t) \Big[ \dot{\psi}(Y_i(t),\tilde\theta_n(t))-\dot{\psi}(Y_i(t),\theta(t)) \Big] + \frac{1}{\sum_{j=1}^n \delta_j(t)} \sum_{i=1}^{n}\delta_i(t)\dot{\psi}(Y_i(t),\theta(t))\\
&=\frac{1}{\sum_{j=1}^n \delta_j(t)/n} \sum_{i=1}^{n} \Big[ \frac{\delta_i(t)}{n}- \frac{b(t)}{n} \Big] \Big[ \dot{\psi}(Y_i(t),\tilde\theta_n(t))-\dot{\psi}(Y_i(t),\theta(t)) \Big] \\
&\qquad \qquad + \frac{b(t)}{\sum_{j=1}^n \delta_j(t)/n} \Big[\frac{1}{n}\sum_{i=1}^{n} \dot{\psi}(Y_i(t),\tilde\theta_n(t))-\dot{\psi}(Y_i(t),\theta(t)) \Big]+ \frac{1}{\sum_{j=1}^n \delta_j(t)} \sum_{i=1}^{n}\delta_i(t)\dot{\psi}(Y_i(t),\theta(t))\\
& \leq \sup_{t \in \rC} \Big\{ \frac{1}{\sum_{j=1}^n \delta_j(t)/n} \Big\} \sup_{t \in \rC} \Big| \sum_{i=1}^{n}\Big[ \frac{\delta_i(t)}{n}- \frac{b(t)}{n} \Big] \Big[ \dot{\psi}(Y_i(t),\tilde\theta_n(t))-\dot{\psi}(Y_i(t),\theta(t)) \Big] \Big| \\
&\qquad \qquad \qquad + \sup_{t \in \rC} \Big\{ \frac{b(t)}{\sum_{j=1}^n \delta_j(t)/n} \Big\} \sup_{t \in \rC} \Big| \frac{1}{n} \sum_{i=1}^{n} \dot{\psi}(Y_i(t),\tilde\theta_n(t))-\dot{\psi}(Y_i(t),\theta(t)) \Big| \\
&\qquad \qquad \qquad + \frac{1}{\sum_{j=1}^n \delta_j(t)} \sum_{i=1}^{n}\delta_i(t)\dot{\psi}(Y_i(t),\theta(t)).\\
\end{aligned}
\end{equation}
As $n \to \infty$, under given conditions and Corollary \ref{cor:entropy_bound}, it is $ O_{P}(n^{-1/2}) + o_{P}(1) + E_{P_Y} \dot\psi(Y(t),\theta(t))$.
By Theorem \ref{thm:fCLT_missing_data_generic}, term (2) converges to Gaussian Process with mean zero and covariance function $\varphi(s,t)$ $=$ $\cov\big\{\psi(Y(t),\theta(t)), \psi(Y(s),\theta(s))\big\}{v(s,t) \over b(s) b(t)}$, where $v(s,t) = E_{P_\delta}[\delta(s) \delta(t)]$. Then it an immediate consequence of Slutsky's lemma.
\end{proof}

\begin{proof} [\underline{Proof of Corollary \ref{cor:ANOVA_test}}]
The convergence of numerator of $T_n$ follows about the same lines as those in the proof of Theorem 1 of \citet*{Shen2004} under functional limit theorem for robust M-estimator under partial sampling structure. The denominator of $T_n$, $trace(\hat\xi(s,t))$, converges to $trace(\xi(s,t))$ with consistent estimator $\hat\xi$. By Slutsky's theorem, proof is done.
\end{proof}

\begin{proof} [\underline{Proof of Corollary \ref{cor:linear_operator}}]
By Karhuneun-Lo\'eve theorem, $\xi(s,t) = \sum_{r=1}^k \kappa_r e_r(t)e_r(s)$ and we have $\sqrt{n} (\hat\theta_n(t)-\theta(t))=\sum_{r=1}^k \eta_r e_r(t)$, where
$$
\eta_r = \int_\rC \sqrt{n}\big(\hat\theta_n(t)-\theta(t)\big) e_r(t) dt \sim AN (0, \kappa_r),~r=1,...,k.
$$
Then we can write
\begin{equation}
\nonumber
\begin{aligned}
\sqrt{n} \langle (\hat\theta_n(\cdot)-\theta(\cdot)), \phi(\cdot) \rangle
&= \sqrt{n} \Big(\langle \hat\theta_n(t), \phi(\cdot) \rangle - c \Big) = \int \big( \sum_{r=1}^k \eta_r e_r(t) \big) \phi(t) dt\\
&= \sum_{r=1}^k \eta_r \int e_r(t) \phi(t) dt= \sum_{r=1}^k \eta_r \langle e_r(\cdot), \phi(\cdot) \rangle,
\end{aligned}
\end{equation}
 Under the assumption of $tr(\xi)<\infty$, $\sum_{r=1}^k \eta_r$ converges in distribution especially to normal distribution. Thus, $\sum_{r=1}^k \eta_r \langle e_r(\cdot), \phi(\cdot) \rangle$ also converges to normal distribution under $\langle e_r(t), \phi(t) \rangle < || e_r(t)|| \cdot || \phi(t)||=c<\infty$. The asymptotic variance is  derived as $\tau^2 = \Var \Big\{ \sum_{r=1}^k \eta_r \langle e_r(\cdot), \phi(\cdot) \rangle \Big\} = \sum_{r=1}^k k_r^2 \kappa_r$, where $k_r^2=\langle e_r(\cdot), \phi(\cdot) \rangle^2 = \int \int_\rC \phi(s) \phi(t) e_r(s) e_r(t) ds dt$.
\end{proof}

\newpage
\section{Appendix: Additional Figure and Results from Simulation Studies }
\label{Simulation}
\subsection{Simulated heavy-tailed Data}
\begin{figure}[t!h!]
  \centering
  \includegraphics[width=6.5in]{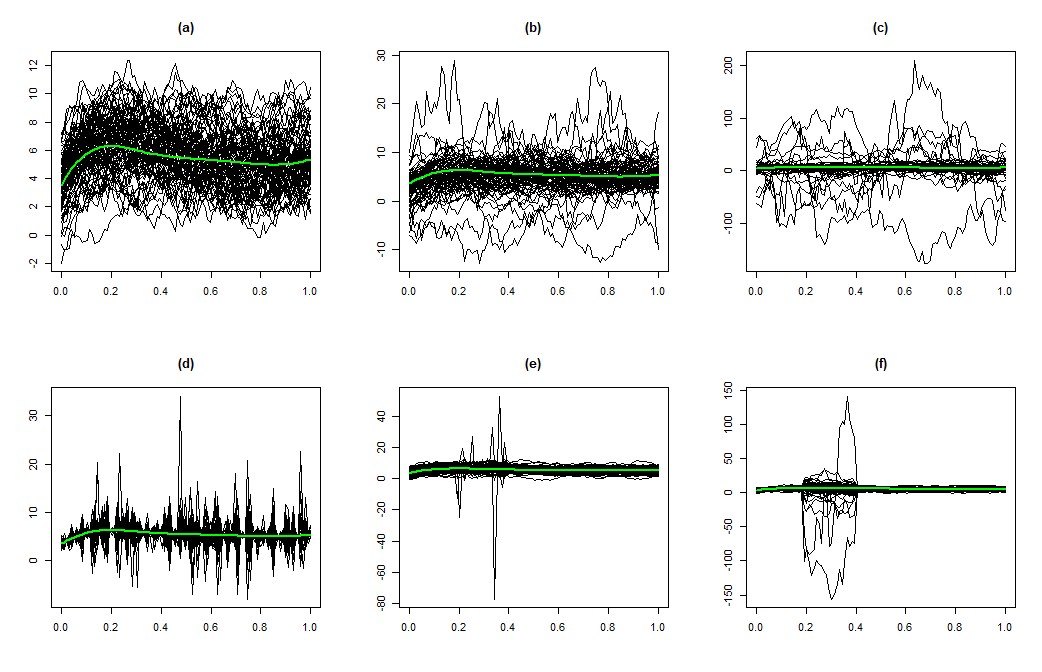}
  \caption{Simulated data from the scenario of (a) Gaussian, (b) $t_3$, (c) Cauchy, (d) white-noise $t_3$ with random scales, Gaussian partially contaminated by (e) Cauchy white-noise, and by (f) Cauchy processes. Green line indicates location function.}\label{fig:outlier}
\end{figure}

We present simulated data in Figure \ref{fig:outlier} under six heavy-tailed or contaminated scenarios considered in Section \ref{sec:sims}.
\pagebreak

\subsection{Additional Results for Simulation Studies}
\begin{figure}[t!h!]
  \centering
  \includegraphics[width=6.5in]{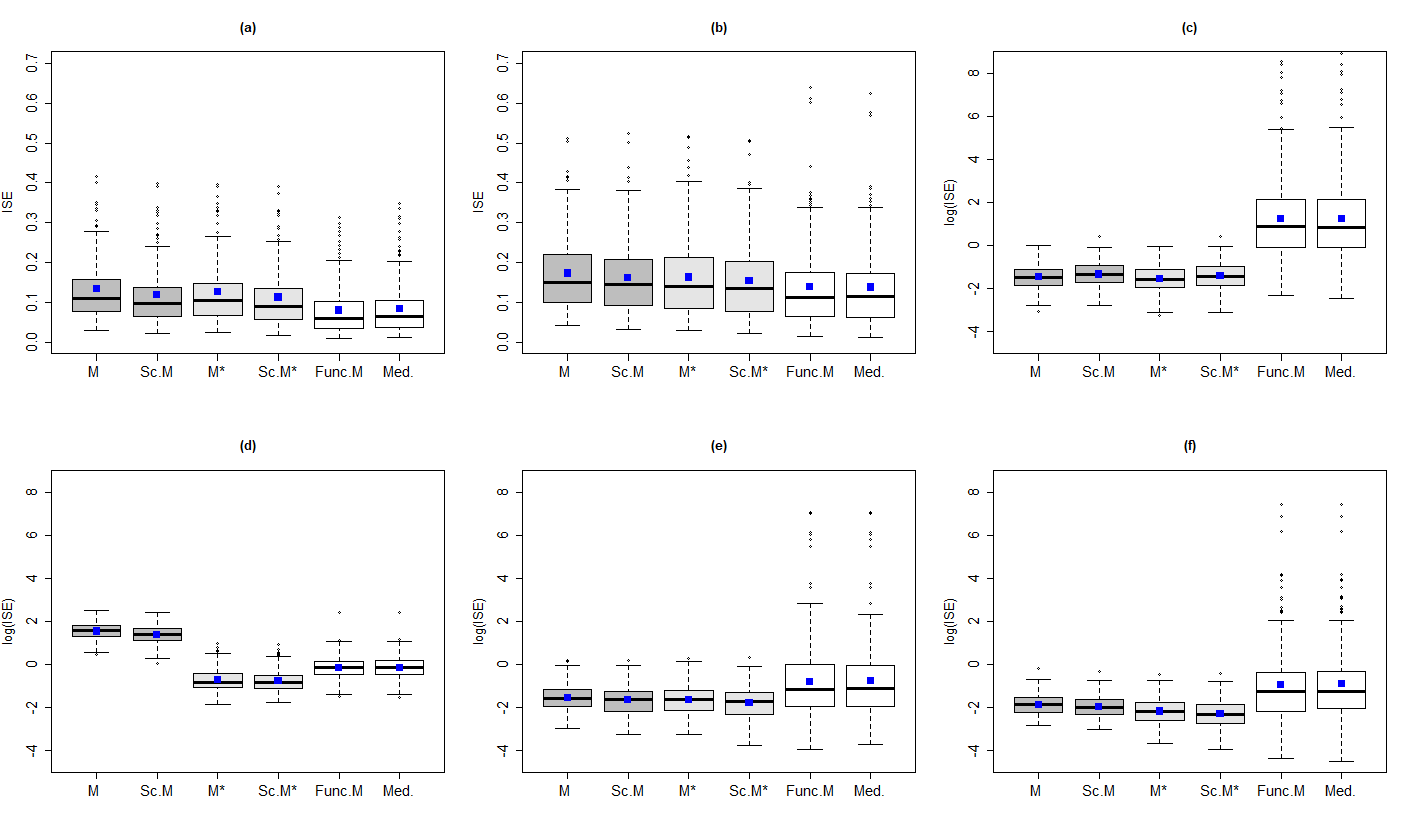}
  \caption{Boxplots of ISE or log transformed ISE over 500 replications from the marginal M-estimator (M), marginal scaled M-estimator (Sc.M), marginal M-estimator under pre-smoothed curves (M*), marginal scaled M-estimator under pre-smoothed curves (Sc.M*), functional M-estimator (Func.M), and functional Median (Med.) under partially observed data at randomly selected interval among a fixed number of pre-specified intervals from (a) Gaussian, (b) $t_3$, (c) Cauchy, (d) white-noise $t_3$ with random scales, Gaussian partially contaminated by (e) Cauchy white-noise, and by (f) Cauchy processes. Blue square dots represent mean values}\label{fig:IMSfixed}
\end{figure}

Figure \ref{fig:IMSfixed} displays the estimation performances of robust estimators under partially observed functional data, especially incomplete curves observed at randomly selected interval among a fixed number of pre-specified intervals. Boxplots show similar behaviors that we observe from the results under random interval structure in Section \ref{sec:sims12}. Our proposed marginal M-estimator achieves superior estimation accuracies compared two competitors under Cauchy data and Gaussian data with Cauchy contaminated noise. Although slightly lower errors are observed from two competing functional estimators under (a) Gaussian and (b) $t(3)$ distributed data compared to ours, differences do not seem significant and almost similar.

Tables \ref{table1}, \ref{table2}, and \ref{table3} provide detailed results for simulations on bootstrapped functional trend test in Section \ref{sec:sims2} with coverage probabilities and the median length of bootstrapped confidence intervals for projected coefficients to quadratic, linear, and constant basis functions, respectively

\pagebreak

\begin{table}[t!]
\caption{Coverage probabilities and the median length of bootstrapped confidence intervals (in parenthesis) of projection coefficient to \underline{Quadratic} basis function from M-estimator (M), scaled M-estimator (Sc.M), and mean over 500 repetitions}
\label{table1}
\myfontsizeAA
\vspace{2mm}
\centering
\begin{tabular} {@{\extracolsep{4pt}}ccccccccccccc@{}} 
\hline
  & \multicolumn{3}{c}{Regular} & \multicolumn{3}{c}{Irregular1}& \multicolumn{3}{c}{Irregular2} & \multicolumn{3}{c}{Dense snippets} \\
  \cline{2-4} \cline{5-7} \cline{8-10} \cline{11-13}\\ [-0.13in]
& Mean & Mt & Sc.M & Mean & M & Sc.Mt & Mean & M & Sc.M & Mean & M & Sc.M\\
\cline{2-4} \cline{5-7} \cline{8-10}\cline{11-13}
\multirow{2}{*}{Gaussian} & 0.94 & 0.942 & 0.942 &0.936 & 0.954& 0.954& 0.942 &0.946 &0.952 &0.910&0.915&0.915   \\
  & (0.14) & (0.15) & (0.15) & (0.23) & (0.24)& (0.24)& (0.22)& (0.24) & (0.24) &(0.17)&(0.19)&(0.19)   \\
 \hline
\multirow{2}{*}{t(3)} & 0.46 & 0.948 & 0.95 &0.94 & 0.95& 0.95&0.938 &0.944 &0.948 &0.930&0.920& 0.920\\
  & (0.23) & (0.18) & (0.18) & (0.35)&(0.29)&(0.29)& (0.35)& (0.28)& (0.28) &(0.27)&(0.22)&(0.22)   \\
 \hline
\multirow{2}{*}{Cauchy} & 0.938 & 0.946 & 0.954 &0.912 &0.950&0.956& 0.902& 0.960& 0.966&0.960&0.880& 0.885 \\
  & (1.55) & (0.22) & (0.24) &(2.00) &(0.37)&(0.42)&(2.41)& (0.37)& (0.40) &(4.10)&(0.29)&(0.31)   \\
 \hline
\multirow{2}{*}{Cont.1} & 0.934 & 0.952 & 0.954 & 0.910& 0.944& 0.946 &0.926 & 0.948& 0.954 &0.920&0.925&0.935 \\
  & (0.83) & (0.21) & (0.22) & (0.95) &(0.34)& (0.35)&(0.86)&(0.34)& (0.35)   &(1.19)&(0.28)&(0.28) \\
\hline
 \multirow{2}{*}{Cont.2} & 0.880 & 0.940 & 0.940 & 0.888& 0.938& 0.940&0.928& 0.950 & 0.954&0.925&0.935& 0.920 \\
  & (0.83) & (0.21) & (0.22) &(0.94) &(0.34)&(0.36)&(0.99)&(0.35)& (0.36)   &(1.17)&(0.28)&(0.28) \\
\hline
\end{tabular}
\end{table}
\pagebreak
\begin{table}[t!]
\caption{Coverage probabilities and the median length of bootstrapped confidence intervals (in parenthesis) of projection coefficient to \underline{Linear} basis function from M-estimator (M), scaled M-estimator (Sc.M), and mean over 500 repetitions}
\label{table2}
\myfontsizeAA
\centering
\vspace{2mm}
\begin{tabular} {@{\extracolsep{4pt}}ccccccccccccc@{}} 
\hline
  & \multicolumn{3}{c}{Regular} & \multicolumn{3}{c}{Irregular1}& \multicolumn{3}{c}{Irregular2}& \multicolumn{3}{c}{Dense snippets}\\
  \cline{2-4} \cline{5-7} \cline{8-10} \cline{11-13}\\ [-0.13in]
& Mean & Mt & Sc.M & Mean & M & Sc.Mt & Mean & M & Sc.M  & Mean & M & Sc.M\\
\cline{2-4} \cline{5-7} \cline{8-10} \cline{11-13}
\multirow{2}{*}{Gaussian} & 0.940 & 0.946 & 0.948 &0.940 & 0.940&0.950 & 0.944 &0.958 &0.960 &0.825& 0.710& 0.730 \\
  & (0.19) & (0.20) & (0.20) & (0.28)&(0.30)&(0.30)& (0.30)& (0.32) & (0.32) &(0.32)&(0.36)&(0.36)   \\
 \hline
\multirow{2}{*}{t(3)} & 0.928 & 0.934 & 0.934 &0.946 &0.958 & 0.966&0.940 &0.946 &0.950&0.820&0.790& 0.815  \\
  & (0.30) & (0.23) & (0.23) &(0.44) &(0.35)&(0.35)& (0.47)& (0.36)& (0.37) &(0.44)&(0.40)&(0.41)   \\
 \hline
\multirow{2}{*}{Cauchy} & 0.908 & 0.948 & 0.942 & 0.896& 0.946& 0.95& 0.902& 0.952& 0.940&0.910&0.680&0.70  \\
  & (2.06) & (0.29) & (0.31) &(2.56) &(0.46)&(0.51)&(3.23)& (0.48)& (0.54) &(3.31)&(0.57)&(0.62)   \\
 \hline
\multirow{2}{*}{Cont.1} & 0.892 & 0.940 & 0.938 & 0.928&0.956&0.954&0.926 & 0.942& 0.950 &0.855& 0.785& 0.805 \\
  & (1.42) & (0.31) & (0.32) &(1.53) &(0.47)&(0.49)&(1.39)&(0.52)& (0.54) &(1.65)&(0.56)&(0.58)   \\
\hline
 \multirow{2}{*}{Cont.2} & 0.918 & 0.942 & 0.944 & 0.888& 0.930& 0.934&0.936& 0.948 & 0.952&0.870& 0.715& 0.725  \\
  & (1.42) & (0.31) & (0.32) &(1.55) &(0.47)&(0.49)&(1.55)&(0.47)& (0.48)   &(1.40)&(0.53)&(0.55) \\
\hline
\end{tabular}
\end{table}
\begin{table}[t!]
\caption{Coverage probabilities and the median length of bootstrapped confidence intervals (in parenthesis) of projection coefficient to \underline{Constant} basis function from M-estimator (M), scaled M-estimator (Sc.M), and mean over 500 repetitions}
\label{table3}
\myfontsizeAA
\vspace{2mm}
\centering
\begin{tabular} {@{\extracolsep{4pt}}ccccccccccccc@{}} 
\hline
  & \multicolumn{3}{c}{Regular} & \multicolumn{3}{c}{Irregular1}& \multicolumn{3}{c}{Irregular2}& \multicolumn{3}{c}{Dense snippets}\\
  \cline{2-4} \cline{5-7} \cline{8-10} \cline{11-13}\\ [-0.13in]
& Mean & Mt & Sc.M & Mean & M & Sc.Mt & Mean & M & Sc.M & Mean & M & Sc.M \\
\cline{2-4} \cline{5-7} \cline{8-10} \cline{11-13}
\multirow{2}{*}{Gaussian} & 0.950 & 0.956 & 0.950 & 0.944& 0.952& 0.952& 0.954 &0.956 &0.954 &0.690& 0.640& 0.665 \\
  & (0.26) & (0.27) & (0.27) &(0.35) &(0.36)&(0.36)& (0.34)& (0.35) & (0.35)&(0.32)&(0.39)&(0.40)    \\
 \hline
\multirow{2}{*}{t(3)} & 0.908 & 0.954 & 0.954 &0.916 &0.934 &0.936 &0.940 &0.958 &0.960 &0.735&0.760& 0.785 \\
  & (0.41) & (0.30) & (0.31) & (0.52)&(0.43)&(0.43)& (0.52)& (0.40)& (0.41)&(0.43)&(0.42)&(0.43)    \\
 \hline
\multirow{2}{*}{Cauchy} & 0.902 & 0.936 & 0.934 & 0.904& 0.956& 0.960& 0.926& 0.954& 0.952&0.915& 0.765& 0.785 \\
   & (2.92) & (0.38) & (0.42) & (3.04) &(0.55)&(0.61)&(3.51)& (0.51)& (0.57) &(3.25)&(0.58)&(0.60)   \\
 \hline
\multirow{2}{*}{Cont.1} & 0.918 & 0.964 & 0.966 & 0.912& 0.950& 0.960&0.922 & 0.948& 0.952 &0.755& 0.690& 0.710 \\
  & (1.18) & (0.34) & (0.34) &(1.25) &(0.47)&(0.48)&(1.17)&(0.49)& (0.51) &(0.99)&(0.54)&(0.55)\\
\hline
 \multirow{2}{*}{Cont.2} & 0.918 & 0.960 & 0.962 & 0.874& 0.930& 0.930 &0.914& 0.942 & 0.940 &0.735& 0.645& 0.680 \\
  & (1.17) & (0.34) & (0.34) &(1.27) &(0.47)&(0.48)&(1.28)&(0.44)& (0.45)&(0.84)&(0.49)&(0.50)    \\
\hline
\end{tabular}
\end{table}
\pagebreak

\subsection{Validity of Robust Inference for Partially Observed Functional Data under Sparse Design Points}
To evaluate the numerical feasibility and performance of robust inference under partially observed functional data recorded at sparse points, we apply sparse sampling scheme to sets of curves generated under five distributional assumptions considered in Section \ref{sec:sims2}.

Specifically, we first define $\epsilon$-equispaced grid points, $t_0, t_1,...,t_{1/\epsilon}$, over $[0,1]$, for sufficiently small $\epsilon >0$, then generate $l_i = \min(v_{i1}, v_{i2})$ and $u_i = \max(v_{i1}, v_{i2})$ from $v_{ij}$, $j=1,2$, i.i.d. from a discrete uniform random variable V on $\{ t_0, t_1,...,t_{1/\epsilon}\}$ to set the lower and upper bounds of the subinterval of each curve. Let $t_{ij},$ $j=1,...,n_i$ denote grid points within each individual random subinterval and $t_{i1}=l_i$, $t_{in_i}=u_i$ by definition. We then assume Bernoulli distribution to draw binary indicator, $\delta(t_{ij}) \overset{i.i.d.}{\sim} Bernoulli(p)$, where $p$ controls the sparsity of the data. In our simulation, we set $p=0.4$ and the sample size as $n=200$.

\begin{figure}[t!]
  \centering
  \includegraphics[width=6.5in]{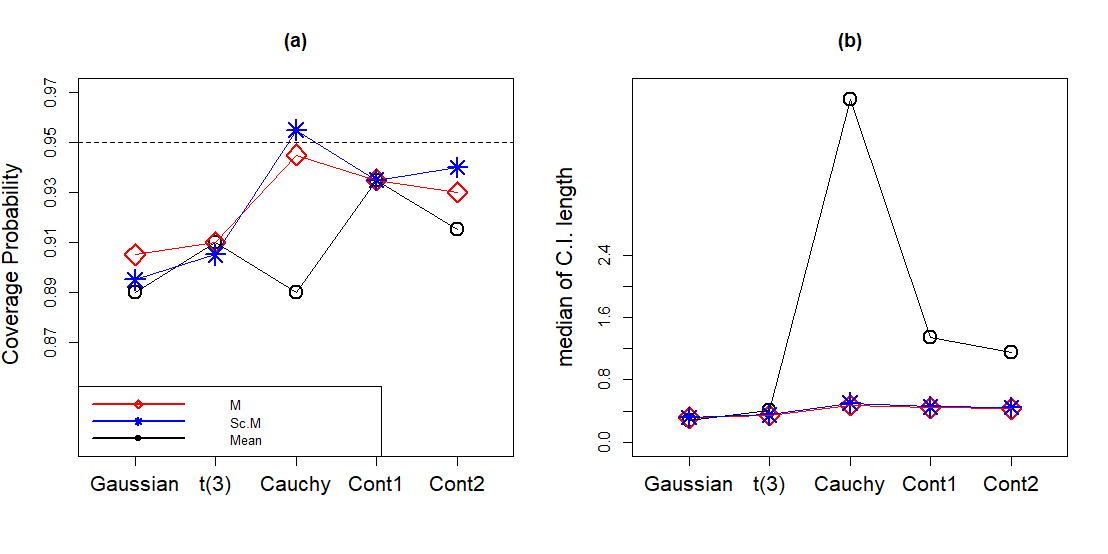}
  \caption{(a) Coverage probabilities of bootstrapped confidence intervals of projection coefficients to quadratic function under Gaussian, $t_3$, Cauchy, and two contaminated data from M-estimator (M), scaled M-estimator (Sc.M), and Mean functions over 500 repetitions under sparse data. (b) Median length of bootstrapped confidence intervals of projection coefficient}\label{fig:projectionSparse}
\end{figure}

As the generated data have regular sparse design points, the marginal M-estimator is applied to each $\epsilon$-equispaced grid point and Figure \ref{fig:projectionSparse} displays coverage probabilities of $95\%$ bootstrapped confidence intervals for quadratic coefficient of location parameter and median length of bootstrapped confidence intervals. First it can be seen that robust inferential test performs well even under sparse design by detecting true quadratic trend with $90-95\%$ coverage probabilities under various distributional settings. We also observe stable behaviors of confidence intervals with almost constant length of confidence intervals even with existence of heavy-tailed curves or contaminations. On the other hand, unstable performance is observed from inferential test based on functional means.

Via simulation studies, we illustrate the numerical feasibility and validity of robust inferential method for fragmented data observed at sparse grid points. Although further extension on theory, especially for conditions on partial sampling process, is required to fit spares structure to our proposed framework, promising numerical results shine a light on the generalization of our approach even to sparsely observed data.

\pagebreak

\par
\bibhang=1.7pc
\bibsep=2pt
\fontsize{9}{14pt plus.8pt minus .6pt}\selectfont
\renewcommand\bibname{\large \bf References }
\expandafter\ifx\csname
natexlab\endcsname\relax\def\natexlab#1{#1}\fi
\expandafter\ifx\csname url\endcsname\relax
  \def\url#1{\texttt{#1}}\fi
\expandafter\ifx\csname urlprefix\endcsname\relax\def\urlprefix{URL}\fi

 \bibliographystyle{chicago}      
  \bibliography{fcltrefs}   

\end{document}